\pdfoutput=1

\documentclass[aip,jcp,reprint,groupedaddress,footinbib]{revtex4-1}
\usepackage{graphicx} %
\usepackage{amsmath}
\usepackage{mathtools} %
\usepackage{bm}

\usepackage{hyperref}
\hypersetup{colorlinks=true,linkcolor=blue,citecolor=blue,urlcolor=blue}

\usepackage{braket} %
\newcommand{\ketbra}[2]{\ket{#1}\!\bra{#2}}

\newcommand{\eu}{\mathrm{e}^}
\newcommand{\iu}{\ensuremath{\mathrm{i}}}
\newcommand{\rmd}{\mathrm{d}}
\newcommand{\half}{\ensuremath{\tfrac{1}{2}}}

\newcommand{\op}[1]{\ensuremath{\hat{#1}}}

\renewcommand{\mathbf}[1]{\bm{#1}}
\DeclareMathOperator{\Tr}{Tr}
\DeclareMathOperator{\tr}{tr}

\DeclareMathOperator{\Imag}{Im}
\DeclareMathOperator{\Erf}{Erf}

\newcommand{\ImF}{\mbox{$\Imag F$}}

\newcommand{\eqn}[1]{Eq.\,(\ref{#1})}

\newcommand{\fig}[1]{Fig.\,\ref{fig:#1}}
\newcommand{\secref}[1]{Sec.\,\ref{sec:#1}}
\newcommand{\Ref}[1]{Ref.~\onlinecite{#1}}
\newcommand{\Refs}[1]{Refs.~\onlinecite{#1}} %

\newcommand{\Heaviside}{h}

\begin{document}

\title{Non-oscillatory flux correlation functions for efficient nonadiabatic rate theory} %

\author{Jeremy O. Richardson}
\email{jeremy.richardson@fau.de}
\author{Michael Thoss}
\affiliation{
Institut f{\"u}r Theoretische Physik und Interdisziplin{\"a}res Zentrum f{\"u}r Molekulare Materialien,
Friedrich-Alexander-Universit{\"a}t Erlangen-N{\"u}rnberg (FAU),
Staudtstra{\ss}e 7/B2,
91058 Erlangen, Germany
}

\date{\today}

\begin{abstract}
There is currently much interest in the development of improved trajectory-based methods
for the simulation of nonadiabatic processes in complex systems.
An important goal for such methods is the accurate calculation of the rate constant over a wide range of electronic coupling strengths
and it is often the nonadiabatic, weak-coupling limit, which being far from the Born-Oppenheimer regime,
provides the greatest challenge to current methods.
We show that in this limit there is an inherent sign problem impeding further development
which originates from the use of the usual quantum flux correlation functions,
which can be very oscillatory at short times.
From linear response theory,
we derive a modified flux correlation function for the calculation of nonadiabatic reaction rates,
which still rigorously gives the correct result in the long-time limit regardless of electronic coupling strength,
but unlike the usual formalism is not oscillatory in the weak-coupling regime.
In particular, a trajectory simulation of the modified correlation function is naturally initialized in a region localized about the
crossing of the potential energy surfaces.
In the weak-coupling limit, a simple link can be found between the dynamics initialized from this transition-state region
and an generalized quantum golden-rule transition-state theory,
which is equivalent to Marcus theory in the classical harmonic limit.
This new correlation function formalism thus provides a platform
on which a wide variety of dynamical simulation methods can be built
aiding the development of accurate nonadiabatic rate theories applicable to complex systems.
\end{abstract}


\maketitle

\section{Introduction}

The simulation of electronically nonadiabatic dynamics in complex molecular systems
poses a significant challenge to current techniques in theoretical chemistry.
\cite{Tully2012perspective}
These dynamics describe the motion of nuclei in systems in which the Born-Oppenheimer approximation is not valid.
This could occur for instance at a conical intersection between two electronic states,
\cite{ConicalIntersections1,Levine2007AIMS}
or as we consider here,
near an avoided crossing.
\cite{[{For an overview, see for instance }] [{ and references therein.}] Stock2005nonadiabatic}
In the latter case, the nonadiabatic limit is of special interest,
in which the electronic coupling, here referred to as $\Delta$, is weak.

A wide variety of methods have been proposed 
for performing nonadiabatic dynamics either numerically exactly or with varying degrees of approximation.
\cite{Stock2005nonadiabatic}
However, the only systems amenable to exact calculations have either very few degrees of freedom
or are model Hamiltonians
such as the spin-boson model of electron transfer in the condensed phase,
which is defined as a two-level system coupled linearly to a bath of harmonic oscillators. \cite{Garg1985spinboson,*Leggett1987spinboson}
Although an exact closed-form expression for the rate constant of this system is not known,
approximate expressions have been formulated by assuming the separability of a reaction coordinate
which bridge the various regimes, \cite{Rips1995ET}
and otherwise difficult numerical calculations can be made practical by exploiting its simple harmonic form. \cite{Weiss}

The treatment of discrete electronic states is naturally described
by exact quantum dynamical methods,
which differ very little from their Born-Oppenheimer equivalents.
Such methods, which have been applied for instance to the spin-boson model,
include hybrid approaches which split the system into a small quantum core and large classical reservoir, \cite{Wang2001hybrid,*Thoss2001hybrid}
and the multilayer multiconfigurational time-dependent Hartree (MCTDH) method, \cite{Wang2003MLMCTDH,*Thoss2006MLMCTDH,Wang2006flux}
which treats all degrees of freedom quantum mechanically.
It is also possible to utilize real-time path-integral approaches
\cite{Mak1991spinboson,Topaler1996nonadiabatic,Muehlbacher2003spinboson,*Muehlbacher2004asymmetric}
to compute the dynamics of system-bath models,
where the influence of the harmonic bath can be formally integrated out analytically.

It is not always necessary to calculate the time evolution of the system explicitly
in order to obtain the nonadiabatic rate constant.
An important example is the golden-rule approach
\cite{*[{See for instance }] [{ for a clear derivation.}] Zwanzig}
which employs an approximation valid only in the $\Delta\rightarrow0$ limit
and gives a rate proportional to $\Delta^2$.
The standard formula, given in \eqn{goldenrule}, 
relies on the complete solution of reactant and product states
and is thus not directly applicable to complex systems,
by which we mean systems for which this set of states are not computable.
Application to systems of interest is only possible after
an approximation is made
that the potential energy surfaces are harmonic with normal modes independent of electronic state.
\cite{Lee2013goldenrule}
A quantum golden-rule formula has been derived in the case of the spin-boson model, \cite{Weiss}
which can be written in closed form only after applying an approximate stationary-phase integral. \cite{Bader1990golden}
The celebrated theory of Marcus \cite{Marcus1956ET}
for the rate of electron transfer in the condensed phase \cite{*[{See for instance }] [{}] ChandlerET}
can be derived from this result in the classical limit
and is therefore also known as the classical golden-rule rate.
One of the most significant predictions of the theory
was the ``inverted regime'', in which the rate of electron transfer in strongly exothermic systems decreases with increasing exothermicity,
and which was later confirmed by experiment.\cite{Miller1984inverted}
We note that the Marcus theory rate is exact in the golden-rule limit for a spin-boson system with classical nuclei,
i.e.\ with heavy masses, %
and thus provides a benchmark against which nonadiabatic dynamical methods can be tested in this limit.

For complex systems,
a procedure commonly followed is to write the golden-rule rate as a time-correlation function
of the energy gap between the diabatic states.
This is then approximately computed using classical trajectories
which evolve on a single surface given by an interpolated average of the diabatic potentials
and should thus not be considered as a simulation of the true dynamics
but rather as a sampling procedure for estimating the rate constant.
The results are not unique due to a dependence of the choice of propagating Hamiltonian
and an associated quantum correction factor.
\cite{Egorov1999goldenrule,Shi2004goldenrule}

A generalization of the golden-rule approach
to compute the rate approximately using statistical mechanics,
instead of explicitly solving for the states or performing time-propagations,
was proposed by Wolynes \cite{Wolynes1987nonadiabatic}
using a semiclassical approximation to an analytic continuation of the flux-flux correlation function.
This gave a nonadiabatic free-energy approach
which can be computed numerically using imaginary-time path integral Monte Carlo
and is therefore applicable to complex systems.
The formalism was rederived by Cao and Voth \cite{Cao1997nonadiabatic}
from a nonadiabatic generalization of the \ImF\ method \cite{Coleman1977ImF,*Affleck1981ImF}
and has been employed in a number of studies of electron transfer in the condensed phase.
\cite{Zheng1989ET,*Zheng1991ET,Bader1990golden,*Marchi1991tunnelling}
The method requires locating a stationary point of the free energy
as a function of imaginary time $\beta'\hbar$
about which a steepest-descent imaginary-time integration is performed.
There would however be a problem with this approach in the Marcus inverted regime where the stationary point
falls outside the interval $[0,\beta\hbar]$ and thus has no meaning for path integrals of period $\beta\hbar$.
We also note that due to the steepest-descent time integration,
the result is not in general exactly equal to the quantum golden-rule rate,
although it may be a good approximation in many cases outside the inverted regime.

A nonadiabatic theory
based on imaginary-time path-integral sampling
has also been formulated\cite{Schwieters1999diabatic}
which attempts to give the rate constant over the whole range of the electronic coupling $\Delta$.
However, in order to ensure the correct $\Delta^2$ behaviour in the golden-rule limit,
an \textit{ad hoc} modification of the barrier height was made.
The method is not valid for biased systems,
and even in the classical limit, does not strictly reduce to the Marcus theory rate for a symmetric system.

In this work, we present an alternative formulation of the golden-rule rate
in terms of statistical mechanical quantities
which differs from previous approaches
and may be a good starting point for the development of a new path-integral golden-rule method
applicable to complex systems.
We show that the new formulation reduces to the Marcus theory rate constant in the classical limit
even in the strongly biased inverted regime.
We argue that it is a type of transition-state theory (TST) and show how it is related to dynamical methods.

We now turn to approximate dynamical methods
based on nonadiabatic trajectory simulations,
which
are intended to be interpreted as real-time dynamics.
An approximate nonadiabatic rate constant can formally be defined in terms of correlation functions computed using the dynamics. \cite{Miller1983rate}
Such approaches are naturally applicable to complex systems,
but an accurate description of the coupling between nuclear and electronic degrees of freedom
causes a significant challenge to theory.
\cite{Stock2005nonadiabatic}

The difficulty posed for simulation of nonadiabatic dynamics
can be clearly understood by contrasting with
the conceptually straightforward case of adiabatic dynamics on a single Born-Oppenheimer surface.
In this case, classical Newtonian dynamics
are derived from a rigorous heavy-mass limit of quantum mechanics
allowing the system to be treated by a molecular dynamics trajectory simulation,
which is far less computationally demanding than approaches based on nuclear wave functions
and also has the advantage that it can be combined with an on-the-fly calculation of the potential.
Due to the inherently quantum nature of the discrete electronic states,
no such classical limit exists for nonadiabatic dynamics %
even in this deceptively simple heavy-mass limit.

The most commonly employed trajectory method is the surface hopping approach of Tully \cite{Tully1990hopping}
in which nuclei evolve classically on one of the adiabatic potential energy surfaces
with hops between states performed randomly according to a particular algorithm.
Many forms exist to determine when the hops should take place and to deal with the energy conservation problem
that arises upon instantaneously changing the potential energy of the system.
A significant failure of the standard implementation \cite{Tully1990hopping} in the context of this work
is that rates do not obey the correct $\Delta^2$ dependence in the golden-rule limit.
\cite{Landry2011hopping,*Landry2012hopping}
However, the method is simple to perform and to couple with on-the-fly electronic-structure calculations
such that it has gained a high popularity.

An alternative approach for describing electronic-nuclear coupling employs
an exact mapping of the Hamiltonian from discrete electronic states
to continuous degrees of freedom.
\cite{Meyer1979nonadiabatic,Stock1997mapping,*Thoss1999mapping}
Approximations have been taken to treat the nonadiabatic dynamics from the resulting Hamiltonian
classically, \cite{Mueller1998mapping,*Mueller1999pyrazine}
semiclassically, \cite{Stock1997mapping,*Thoss1999mapping,Sun1997mapping,Bonella2003mapping,Miller2009mapping}
using the linearized semiclassical approach, \cite{Sun1998mapping,Wang1999mapping}
or with centroid-molecular dynamics. \cite{Liao2002mappingCMD}
Recently other methods based on the mapping approach have appeared
for treating thermal initial states, \cite{Ananth2010mapping}
using a ring-polymer molecular dynamics (RPMD) Hamiltonian, \cite{mapping,Ananth2013MVRPMD}
or in combination with partially linearized real-time path integrals.\cite{Huo2013PLDM}

Other dynamical approaches
employ mean-field approximations, \cite{Micha1983Ehrenfest}
multiple spawning, \cite{BenNun2002AIMS}
the quantum-classical Liouville equation \cite{Kapral2006QCL}
or an exact factorization of the complete molecular Hamiltonian. \cite{Agostini2013MQC}
Approximate methods intended solely for the estimation of nonadiabatic rate constants
include those which
treat the transferred electron explicitly with RPMD, \cite{Menzeleev2011ET,Kretchmer2013ET,Shushkov2013instanton}
or use a modified RPMD Hamiltonian to enforce the correct $\Delta^2$ dependence in the golden-rule limit. \cite{Menzeleev2014kinetic}

It is a major goal to formulate a nonadiabatic theory which is able to compute the rate constant
over a wide range of coupling strengths, $\Delta$, for complex systems.
It is obvious that
in the adiabatic limit, where the Born-Oppenheimer approximation is valid,
such a theory should
reduce to something equivalent to classical rate theory, classical TST
or ring-polymer TST, \cite{RPMDrefinedRate,rpinst}
which is known to be exact in the absence of recrossing. \cite{Hele2013QTST,*Althorpe2013QTST,Hele2013unique}
In this limit, %
nonadiabatic dynamical approaches \cite{mapping,Ananth2013MVRPMD}
can be applied to compute the small amount of dividing-surface recrossing due to nonadiabatic effects. 
However, this approach breaks down in the weak-coupling limit
where the transmission coefficient is extremely small and therefore inefficient to calculate.
A quite different form of rate theory is required in this limit which we discuss in this work.

First we outline a general derivation of the rate constant from linear response theory in \secref{linearresponse}\@.
We then explain the oscillatory problem in the nonadiabatic limit in \secref{goldenrule}
and suggest a solution based on linear response in \secref{nonoscillatory}\@.
We analyse the short-time limit of our approach and hence its relation to dynamical methods in \secref{dynamics}\@.
This also leads to a new golden-rule TST formulation described in \secref{PIGTST}
and we show that its classical limit gives the Marcus theory rate in \secref{classical}\@.
Finally we summarize our findings in \secref{conclusions} and discuss future directions.

\section{Linear Response Theory}
\label{sec:linearresponse}

Here we review quantum linear response theory
\cite{*[{See for instance }] [{ for a good introduction.}] ChandlerGreen}
from which we derive a general formulation for the rate constant.
Using this result, we shall suggest a new approach based on a modification to the traditional flux correlation functions
which has greatly improved properties in the nonadiabatic limit.

Consider the following chemical reaction
described by first-order rate constants
and occurring in a system defined by the Hamiltonian $H$:
\begin{align}
	A \xrightleftharpoons[k_-]{k} B,
\end{align}
where $k$ is the forward rate constant and $k_-$ is that for the reverse reaction.
We imagine that
the system has been prepared in a nonequilibrium state
with a small perturbation $-\lambda H_1$ applied to the Hamiltonian
such that the Boltzmann operator becomes $\rho_1 = \eu{-\beta(H-\lambda H_1)}$, %
with a reciprocal temperature $\beta=(k_\mathrm{B}T)^{-1}$.
The perturbation is turned off at $t=0$,
and after a short transient time,
the rate constant can be found from the exponential relaxation of the system to equilibrium.

We define projection operators $A$ and $B$ to describe the reactants and products
as defined by experiment
in such a way as to cover the Hilbert space of the system, i.e.\ \mbox{$A+B=1$}.
The assumption is that for long enough times $t>0$,
the phenomenological rate equations
\begin{equation}
	\braket{\dot{B}(t)}_1 = - \braket{\dot{A}(t)}_1 = k\braket{A(t)}_1 - k_-\braket{B(t)}_1
	\label{phenomenological}
\end{equation}
hold,\cite{Chandler1978TST} where for any operator $G$,
\begin{equation}
	\braket{G(t)}_1 = \frac{\Tr\,[\rho_1 G(t)]}{\Tr\,[\rho_1]}.
	\label{Gt}
\end{equation}
The Heisenberg time-dependence of an operator is given by
$G(t)=\eu{\iu Ht/\hbar} \, G \, \eu{-\iu Ht/\hbar}$
and its time derivative by
$\dot{G}\equiv\rmd G/\rmd t=\frac{\iu}{\hbar}[H,G]$,
with the commutator $[H,G]=HG-GH$.
If the relaxation to equilibrium cannot be described by an exponential function
then the chemical reaction under consideration is not a rate process,
the phenomenological equations are not valid,
and thus the rate constant is undefined.
This situation is easily identified
from the absence of a plateau in 
the flux correlation functions discussed in this work.
We will here consider only reactions for which a rate constant can be defined.

Using $\braket{\dot{B}}=0$ and the detailed balance condition at equilibrium, $k \braket{A} = k_- \braket{B}$,
we may rewrite \eqn{phenomenological} as
\begin{align}
	\braket{\Delta\dot{B}(t)}_1
	&= k\braket{\Delta A(t)}_1 - k_-\braket{\Delta B(t)}_1 \\
	&= (k+k_-)\braket{\Delta A(t)}_1,
\end{align}
where $\Delta G=G-\braket{G}$, the unperturbed equilibrium value is $\braket{G}=Z_G/Z$,
the projected partition function is $Z_G=\Tr\,[\eu{-\beta H}G]$
and the total partition function is $Z=\Tr\,[\eu{-\beta H}]$.
Using detailed balance again and rearranging, we obtain
\begin{equation}
	\frac{k}{\braket{B}} = \frac{\braket{\Delta\dot{B}(t)}_1}{\braket{\Delta A(t)}_1}.
\end{equation}
In order to to find a formulation of the rate in terms only of equilibrium quantities,
we consider the first-order effect of the perturbation on \eqn{Gt}:
\begin{align}
	\frac{\rmd}{\rmd\lambda} \braket{G(t)}_1
	&= \beta \,\big( \braket{H_1, G(t)}_1 - \braket{G(t)}_1 \braket{H_1}_1 \big)\, ,
\end{align}
where we introduce the perturbed Kubo-transformed correlation function \cite{KuboBook}
\begin{multline}
	\braket{H_1, G(t)}_1
	= \frac{1}{\Tr\,[\rho_1]\beta}
		\int_0^\beta
		\rmd\beta' \\ \times
		\Tr\left[\eu{-\beta'(H-\lambda H_1)} \, H_1 \, \eu{-(\beta-\beta')(H-\lambda H_1)} \, G(t)\right].
	\label{Kubo1}
\end{multline}
From this, we can write a Taylor series expansion,
which using the equilibrium condition $\braket{\Delta G(t)}=0$, is
\begin{align}
	\braket{\Delta G(t)} = \beta \lambda \braket{H_1, \Delta G(t)} + O(\lambda^2),
\end{align}
and here the equilibrium Kubo correlation function $\braket{H_1, \Delta G(t)}$ is defined as in \eqn{Kubo1} with $\lambda=0$.
It is the truncation of this series to first order which gives us the ``linear response'' to the perturbing Hamiltonian.

The thermal rate constant may therefore be computed with
\begin{align}
	\frac{k}{\braket{B}}
	= \frac{\braket{H_1,\Delta\dot{B}(t)}}{\braket{H_1,\Delta A(t)}}
	= -\frac{\braket{\dot{H}_1, B(t)}}{\braket{H_1,\Delta A(t)}},
	\label{generalKubok}
\end{align}
where we have used the time-symmetry properties of the Kubo-transformed correlation function.
As for all formulae for the rate constant in this work,
we take a value of $t$ in the plateau of the correlation function.
This should be longer than the time taken for the transient behaviour to relax but shorter than the time taken for the system to reach equilibrium.
The most efficient methods are those which minimize the transient time such that the rate constant can be computed after only a short propagation time,
or even, as in the case of TST, without propagation at all.
The transient dynamics are not described by linear response theory
as a consequence of the assumptions inherent in the phenomenological equations,
which are only valid after the transient behaviour has relaxed.

So far we have not specified the form of the perturbation $H_1$
and the rate constant is independent of this choice.
The traditional definition of $H_1=B$ simply perturbs the initial ratio of reactants to products
and gives
\begin{equation}
	\frac{k}{\braket{B}} = - \frac{\braket{\dot{B}, B(t)}}{\braket{B,\Delta A(t)}},
\end{equation}
which for a reactive scattering problem
or in the limit of a slow condensed-phase reaction,
where we neglect $\braket{B,A(t)}$ such that $\braket{B,\Delta A(t)}\approx-\braket{B}\braket{A}$,
\cite{*[{Formulae equivalent to those from }] [{ can be used if this slow-reaction limit is not valid.}] Craig2007condensed}
leads to the familiar flux-side correlation function formalism,
albeit in the Kubo-transformed version,
\begin{subequations}
\label{kfluxcorr}
\begin{equation}
	k = \frac{\braket{F, B(t)}}{\braket{A}},
	\label{kfluxside}
\end{equation}
where the flux $F \equiv \dot{B}$ throughout.
The flux-flux form is found by differentiating the right-hand side with respect to time
and the rate is computed from this as
\begin{equation}
	k = \int_0^t \frac{\braket{F,F(\tau)}}{\braket{A}} \, \rmd \tau.
	\label{kfluxflux}
\end{equation}
\end{subequations}
With the notable exception of the classical Born-Oppenheimer flux-side function which is discontinuous, \cite{Chandler1978TST}
for many purposes the computation of the flux-side is equivalent to the computation of the flux-flux function,
and we refer to the pair collectively as flux correlation functions.

For products defined by a nuclear-configuration dividing surface at $x=0$,
$B$ can be written as the Heaviside step function $\Heaviside[\op{x}]$,
and we recover the result of Yamamoto, \cite{Yamamoto1960rate}
\begin{equation}
	k = \frac{\braket{\tfrac{1}{2m}(\delta[\op{x}]\,\op{p}+\op{p}\,\delta[\op{x}]), \Heaviside[\op{x}](t)}}{\braket{\Heaviside[-\op{x}]}},
	\label{kadiabatic}
\end{equation}
where $\hat{x}$ and $\hat{p}$ are the quantum mechanical operators of position and momentum for a particle of mass $m$.
After substituting the quantum operators for classical variables,
it becomes the formula commonly used in classical rate calculations
either directly or in its TST form,
given by the $t\rightarrow0_+$ limit.
\cite{Chandler1978TST,ChandlerGreen}
Equation~(\ref{kadiabatic}) is the appropriate formulation for a non-diffusive reaction
proceeding either on a single Born-Oppenheimer surface
or in the adiabatic limit of a multi-surface system.

Although the Kubo-transformed flux correlation functions appear naturally from quantum linear response theory, \cite{KuboBook}
there are many generalized functions which differ only at short times
and which could therefore be used to compute the rate constant instead.
In particular, there exists an infinite set of functions with the form
\begin{align}
	C_{\dot{H}_1G}^{\beta'}(t) = \Tr\left[\eu{-\beta' H} \dot{H}_1 \eu{-(\beta-\beta')H} G(t)\right],
\end{align}
where the parameter $\beta'$ takes values between $0$ and $\beta$.
These correlation functions are, in general, complex at short times
except for the symmetrized version $\beta'=\half\beta$
which, like the Kubo-transform,
\begin{align}
	\label{Ckubo}
	\tilde{C}_{\dot{H}_1G}(t) \equiv Z \braket{\dot{H}_1,G(t)} = \frac{1}{\beta}\int_0^\beta C_{\dot{H}_1G}^{\beta'}(t) \, \rmd\beta',
\end{align}
is everywhere real.

We thus come to our most general formulation for the rate,
which is, from \eqn{generalKubok},
\begin{align}
	\frac{k}{\braket{B}}
	= - \frac{C_{\dot{H}_1B}(t)}{C_{H_1\Delta A}(t)},
	\label{generalk}
\end{align}
where $C$ may take the form either of $C^{\beta'}$ with any choice of $\beta'$ or of the Kubo-transform $\tilde{C}$,
and again $t$ is chosen within the plateau.

The special case of $H_1=B$,
within the limit of a slow reaction,
gives the rate in terms of the flux correlation functions $C^{\beta'}_{FB}(t)$ and $C^{\beta'}_{FF}(t)$
in an equivalent way to \eqn{kfluxcorr}.
This formulation was first derived by Miller from scattering theory \cite{Miller1974QTST} and shown \cite{Miller1983rate}
to share the same long-time limit as the Kubo-transformed version given by Yamamoto. \cite{Yamamoto1960rate}
Of the set of possible correlation functions, 
it is the symmetric version which is most commonly computed by exact wave-function \cite{Miller1983rate,Wang2006flux}
or real-time path-integral methods. \cite{Topaler1996nonadiabatic}
The Kubo-transformed version is often the most similar to the correlation functions calculated by classical mechanics, \cite{RPMDcorrelation}
and it is a generalization of the Kubo-transformed flux-side function \cite{Hele2013QTST,*Althorpe2013QTST}
which is approximated by standard Born-Oppenheimer RPMD rate theory.\cite{RPMDrefinedRate}

The flux-correlation-function formulation of Miller \textit{et al.} \cite{Miller1983rate}
has been successfully applied to numerous reactions
and is frequently used as a starting point for further developments in quantum rate theory.
However, as we have shown, a more general formulation, \eqn{generalk}, exists if the nonequilibrium perturbation is chosen such that $H_1\ne B$.
This generalization is similar to that made by
Refs.~\onlinecite{Borkovec1990rate} and \onlinecite{Montero1997rate,*MolSim}
for a classical rate calculation,
in which various new perturbations were suggested to improve the efficiency 
for the case of a diffusive reaction coordinate.

Considering applications to electronically nonadiabatic reactions,
it is widely known that for weak electronic coupling,
the reactants and products are poorly described by the choice of $A$ and $B$
based on a nuclear-configuration dividing surface
and would thus lead
to an inefficient method for computing the rate with very poor statistical sampling.
This can be seen using an approximate result from Landau-Zener theory \cite{Zener1932LZ,*[{See for instance }] [{}] Nitzan}
which states that in this limit the probability of a trajectory being reactive should be proportional to $\Delta^2$.
One cannot therefore assume, as in adiabatic TST,
that trajectories with positive momentum at the dividing surface are reactive and will not recross.
A more appropriate dividing-surface concept is provided by considering projections onto electronic states.
\cite{Topaler1996nonadiabatic,Wang1999mapping,Wang2006flux}
However, with this form, we find strong oscillations in the flux correlation functions in some parameter regimes.
\cite{*[{See }] [{ and its supplemental material.}] Huo2013PLDM}
We discuss this issue in the following section
and offer a solution to the problem
based on choosing a new nonequilibrium perturbation $H_1$ appropriate for dynamics in the nonadiabatic limit.

\section{Flux correlation functions in the nonadiabatic limit}
\label{sec:goldenrule}

We consider now the traditional use of the flux correlation function approach
for the computation of rates in the nonadiabatic limit. \cite{Topaler1996nonadiabatic,Wang1999mapping,Wang2006flux,Huo2013PLDM}
We analyse the properties of these functions
and our proposed modifications
in the strict golden-rule limit,
by which we mean the limit $\Delta\rightarrow0$
such that a Taylor series of the rate constant formula can be well approximated by truncation after the $O(\Delta^2)$ term.
The assumption is that the correlation functions will have similar properties
whenever $\Delta$ is fairly small,
but not necessarily so small that the golden-rule approach is quantitatively correct.
We refer to this regime as the weak-coupling limit,
in which the rate-limiting step is the hop between diabatic surfaces,
as opposed to the surmounting of a potential barrier in coordinate space. %
The system which we consider is very general
and although we perform the analysis with the assumption that all reaction and product states are known,
we shall derive a formulation that requires only the existence of these states and not a complete knowledge of them.
This allows the theory described herein to be applied, in principle,
equally well to large complex systems as to simplified models.

Consider two sets of orthogonal vibronic states, %
representing for instance multidimensional nuclear vibrational wave functions on two diabatic potential energy surfaces.
The vibronic states are defined such that $\ket{\mu}$ is the product of a vibrational state with the diabatic electronic state $\ket{0}$,
and equivalently $\ket{\nu}$ with the diabatic state $\ket{1}$.
They obey the following rules when projected onto the electronic states
$\ket{\mu} = \ket{0}\braket{0|\mu}$, %
$\ket{\nu} = \ket{1}\braket{1|\nu}$,
$\braket{1|\mu}=0$ and
$\braket{0|\nu}=0$,
and hence
$\braket{\mu|\nu}=0$.
We refer the reader to Appendix~\ref{sec:vibronic} for a wave-function representation of our bra-ket notation.

In the absence of electronic coupling, the zero-order Hamiltonian can be written in terms of the energies of these states as
\begin{equation}
	H_0 = \sum_\mu \ket{\mu}E_\mu\bra{\mu} + \sum_\nu \ket{\nu}E_\nu\bra{\nu}.
\end{equation}
With electronic coupling parameter $\Delta$, the total Hamiltonian is
\begin{subequations}
\begin{align}
	H &= H_0 + \Delta \big( \ketbra{0}{1} + \ketbra{1}{0} \big) \\
	&= H_0 + \sum_{\mu\nu} \ket{\mu} \Delta_{\mu\nu} \bra{\nu} + \ket{\nu} \Delta_{\nu\mu} \bra{\mu},
\end{align}
\end{subequations}
where the vibronic coupling matrix is defined as
\begin{align}
	\Delta_{\mu\nu}
	= \Delta_{\nu\mu}^* 
	= \Delta\braket{\mu|0}\braket{1|\nu}.
	\label{couplingmatrix}
\end{align}
Note that even though we have assumed a constant coupling $\Delta$ between the electronic surfaces,
the coupling between vibronic states is necessarily state-dependent.
The extension of these methods
to treat systems with configuration-dependent electronic coupling $\Delta(x)$ would therefore be trivial.

In the case of a continuum of states,
we perform the replacement $\sum_\nu \ketbra{\nu}{\nu} \rightarrow \int\ketbra{\nu}{\nu} \rmd E_\nu$,
ensuring of course the correct normalization of the vibronic states in energy-space.
Only in the limit of a continuum of product states can reflections be avoided
such that the correlation function reaches a plateau and a rigorous rate constant can be defined.
This is the case either in a finite-dimensional system with scattering eigenstates
or in an infinite-dimensional system with bound eigenstates such as those found in the condensed phase. \cite{Weiss}

We could define the projection operators $A$ and $B$ using a dividing surface dependent on the nuclear configuration 
as in \eqn{kadiabatic} and \Ref{Schwieters1999diabatic}.
This is appropriate in the adiabatic, strong-coupling limit where the nuclear dynamics are similar to the motion
on a single mean-field potential energy surface.
However, as pointed out above, in the nonadiabatic, weak-coupling limit,
a better choice for the projection onto reactant states is $A=\ketbra{0}{0} = \sum_\mu\ketbra{\mu}{\mu}$,
and onto product states $B=\ketbra{1}{1} = \sum_\nu\ketbra{\nu}{\nu}$.
The two approaches often give almost identical rates for symmetric systems with strong vibronic coupling
but can be completely different if the equilibrium nuclear distributions on each diabatic state significantly overlap.
\footnote{As an extreme example, consider the case of a system where the two diabatic potentials have a minimum at the same nuclear configuration.
A dividing surface would not then be able to differentiate between products and reactants.}
This is not a failure of the computational methods used but is a consequence of how the rate constant
is defined by the phenomenological equations.
It is therefore important
to choose the approach which is equivalent to the experiment or thought experiment that the theory is attempting to reproduce.
\footnote{In fact, because of this it is not obvious if it is possible to find a single unified formula to give the rate constant
for any value of $\Delta$ from the adiabatic to the nonadiabatic limit.
There is no smooth connection as the very meaning and definition of the rate constant is different in these two regimes.
}
In this work we are considering the nonadiabatic limit and thus take the latter approach,
in which case the flux operator is defined as 
\begin{subequations}
\begin{align}
	F \equiv \dot{B}
	&= \frac{\iu}{\hbar} \Delta \big( \ketbra{0}{1} - \ketbra{1}{0} \big) \label{flux} \\
	&= \frac{\iu}{\hbar} \sum_{\mu\nu} \ket{\mu} \Delta_{\mu\nu} \bra{\nu} - \ket{\nu} \Delta_{\nu\mu} \bra{\mu}.
\end{align}
\end{subequations}

Using perturbation theory to solve for the long-time dynamics
in the $\Delta\rightarrow0$ limit
of the system
initialized by a thermal distribution of reactant states,
one finds that 
the rate is given by\cite{Zwanzig}
\begin{subequations}
\label{goldenrule}
\begin{align}
	k &= \frac{2\pi}{\hbar} \sum_{\mu} \frac{\eu{-\beta E_\mu}}{Z_A} \int |\Delta_{\mu\nu}|^2 \delta(E_\mu-E_\nu) \, \rmd E_\nu \\
	&= \frac{2\pi}{\hbar} \sum_{\mu} \frac{\eu{-\beta E_\mu}}{Z_A} |\Delta_{\mu\bar{\nu}}|^2, %
\end{align}
\end{subequations}
where here $Z_A=\sum_\mu \eu{-\beta E_\mu}$
and $\ket{\bar\nu}$ is the product state (or sum over degenerate states) with energy $E_{\bar\nu}=E_\mu$.
This is the standard form of the golden-rule thermal rate expression,
given as a Boltzmann-weighted sum over of the reactant states of the system.
In \secref{PIGTST}, we will derive an equivalent expression using the quantum trace,
which can, in principle, be calculated in any basis.

Other rate theories \cite{Wolynes1987nonadiabatic,Topaler1996nonadiabatic,Wang1999mapping,Wang2006flux,Huo2013PLDM}
are typically based on the traditional flux correlation formalism, $C^{\beta'}_{FB}(t)$ and $C^{\beta'}_{FF}(t)$,
using these reactant and product projection operators.
\footnote{In this case, the operator $B$ is the projection onto a diabatic state and not a side operator at all.
Nonetheless we retain the familiar terminology for the flux-side correlation function.}
In the language of \secref{linearresponse},
this is equivalent to a derivation from linear response theory with the nonequilibrium perturbation $H_1=B\equiv\ketbra{1}{1}$
and can lead to highly oscillatory functions.
However, in the next section, we shall show that a better choice exists from which the rate constant is more easily defined.

\begin{figure}
	\includegraphics{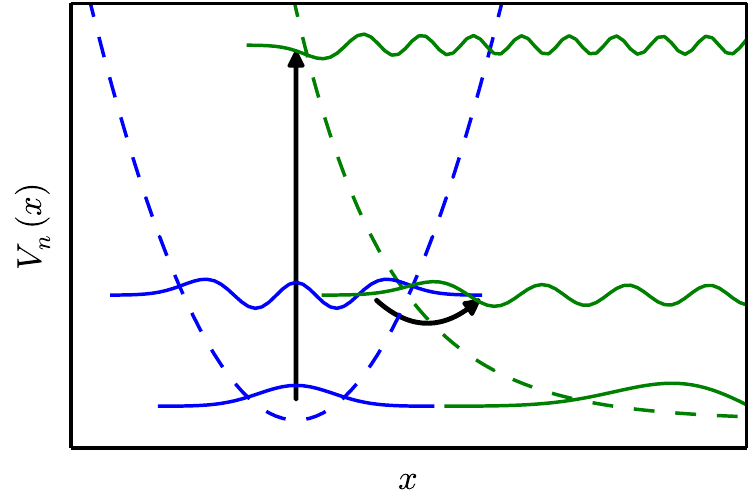}
	\caption{The model system potential energy surfaces (dashed lines) for diabatic states $\ket{n}=\ket{0}$ (blue) and $\ket{1}$ (green)
	with some representative vibronic states indicated by the real parts of their wave functions plotted with arbitrary amplitude at the corresponding energy.
	Two possible transitions from reactant to product states are indicated by black arrows,
	the curved one representing an energy-conserving transition, whereas the straight arrow represents a spurious transition.
	Note that as these transitions are not coupled with absorption or emission of radiation,
	only energy-conserving transitions contribute to the rate constant.
	}
	\label{fig:system}
\end{figure}

To illustrate the problems of the traditional method
and the improvement made by the solution we shall propose,
we introduce the nonadiabatic equivalent of the
standard one-dimensional escape from a metastable well problem, \cite{Haenggi1990rate}
which could describe for instance the dissociation of a molecular species via electron transfer often encountered in the context of predissociation.
This allows us to consider a continuum of product states in a one-dimensional problem where exact results are available.
The model Hamiltonian is defined as
\begin{subequations}
\label{modelH}
\begin{gather}
	H = \frac{\op{p}^2}{2m} + V(\op{x}),
\intertext{with}
	V(\op{x}) = V_0(\op{x})\ketbra{0}{0} + V_1(\op{x})\ketbra{1}{1} + \Delta\big(\ketbra{0}{1} + \ketbra{1}{0}\big),
	\label{Vmatrix}
\end{gather}
\end{subequations}
and diabatic potential energy surfaces
\begin{align}
	V_0(x) &= \half m \omega^2 x^2 \label{Harmonic} \\
	V_1(x) &= D \exp[-2\alpha(x-x_0)] - \epsilon.
	\label{Morse}
\end{align}
Unless otherwise stated, we consider an unbiased system with exothermicity $\epsilon=0$,
a mass of $m=1$
and parameters given by
$\omega=1$, $D=2$, $\alpha=0.2$, $x_0=5$,
at a temperature such that $\beta=3$
and with weak electronic coupling independent of nuclear position $\Delta=0.01$.
We use reduced units such that $\hbar=1$ and energies are effectively measured in units of $\omega$.
The reactant and product nuclear wave functions
needed for the numerical calculation of the coupling matrix via quadrature
are given in Appendix~\ref{sec:wavefunctions}.
The potential surfaces with some representative vibronic states can be seen in \fig{system}.

We note that this model is merely used to illustrate the problem of oscillatory correlation functions
and the improvements made by the introduction of our modified flux-side correlation function formalism in the next section.
An infinite-dimensional condensed-phase problem will also have a continuum of product states, \cite{Weiss}
and as already mentioned, it is simple to extend our analysis to include a nuclear-dependent coupling $\Delta(x)$.
Our findings should therefore apply equally well
to electron transfer in multidimensional complex systems.

\begin{figure}
	\includegraphics{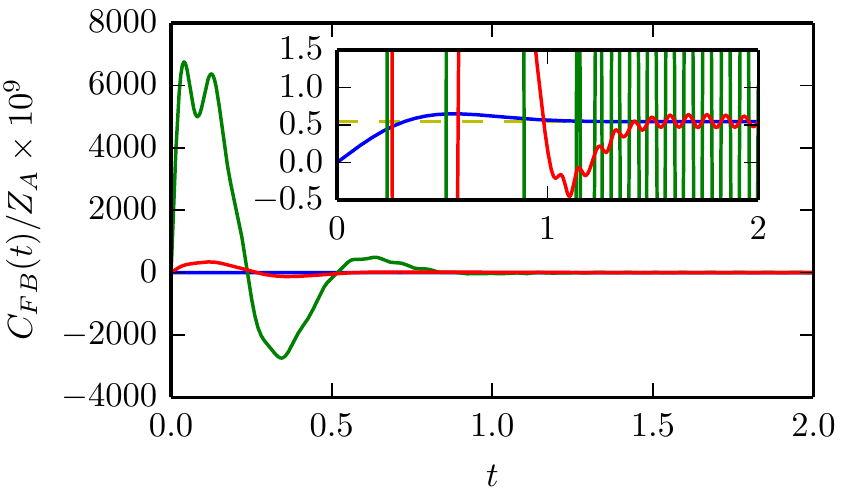}
	\caption{Three flux-side correlation functions for the model system:
	the symmetric version with $\beta'=\tfrac{1}{2}\beta$ is shown in blue,
	the real part of the standard version with $\beta'=0$ in green,
	and the Kubo transform in red.
	All three converge on the golden-rule rate (yellow dashed) in the long time limit
	as can be seen in some of the cases in the magnified inset.
	}
	\label{fig:Kubo}
\end{figure}

We display the time-dependence of various flux-side correlation functions computed for the model system in \fig{Kubo}.
In the numerical calculations,
we took a discrete set of product states and
employed a cut-off for the highest-energy state considered.
The Hamiltonian in the resulting finite basis was diagonalized to give the eigenstates
from which the correlation function is calculated.
The product-state density and cut-off were then increased until convergence of the results was achieved.

It is seen in the case of the unbiased system 
that at short times, the standard correlation function is extremely oscillatory with a large amplitude
and plateaus to the correct rate constant only after a very long transient time (not shown in the figure).
The symmetric version, on the other hand, is quite smooth and tends much sooner to its long-time limit.
The Kubo transformed correlation function, which is an average of such correlation functions as these,
is dominated by its integrand in the limits $\beta'\rightarrow0$ and $\beta'\rightarrow\beta$
and, although totally real, is also strongly oscillatory at short times.
We will describe functions as oscillatory if they at any time considerably overshoot their long-time limit.

Any numerical sampling procedure attempting to compute the rate constant from the Kubo-transformed or standard correlation functions
will have significant numerical convergence problems.
It is unfortunately exactly this type of correlation function that one would like to compute with trajectory simulations
including nonadiabatic RPMD. \cite{mapping}
Many previous approaches have concentrated on computing the smoother symmetric correlation function,
whether based on trajectories employing mapping variables, \cite{Wang1999mapping}
real-time path integrals \cite{Topaler1996nonadiabatic} or MCTDH. \cite{Wang2006flux}
We note, however, that for lower temperatures or strongly biased systems, even the symmetric correlation function can become oscillatory
and could cause problems for these methods as well.
The onset of this regime is observed in the biased systems
considered in \Ref{Wang2006flux} and the real-time path-integral calculations of \Ref{Menzeleev2011ET}
and is discussed in \Ref{Huo2013PLDM}.

What is needed is a new correlation function formalism
from which the rate constant can rigorously be extracted from its long-time limit
but which is not oscillatory at short times.
We address this in the next section.

\section{Non-oscillatory correlation functions}
\label{sec:nonoscillatory}

The causes of the oscillations in the flux-side correlation function
are transitions between low-energy reactant states and high-energy product states
which have a significant overlap in nuclear-configuration space.
Of course, these spurious transitions have different phases and cancel out in the long-time limit
leaving only energy-conserving transitions occurring between degenerate states,
hence the appearance of the delta function in the golden-rule formula, \eqn{goldenrule}.
A schematic of transitions contributing to the short-time limit is given in \fig{system}.
From the viewpoint of a trajectory-based method,
the short-time limit would be dominated by trajectories 
hopping vertically from configurations deep in the reactant or product wells
and not at the crossing of the potentials where one would expect reactive pathways to be located.
It is well-known to be an ill-posed problem to cancel phases using numerical sampling methods,
and thus the long-time limit would be very difficult to converge.

Rather than continuing to use the traditional flux-side formalism,
we return to the general formula for the rate constant derived by linear response theory in \eqn{generalk}
but retain the definitions of $A$ and $B$ from the previous section as projections onto the diabatic surfaces.
We suggest the following modified nonequilibrium perturbation to avoid the oscillation problems at short times,
\begin{subequations}
	\label{H1} 
\begin{align}
	H_1 &= \iint \delta(H - E) \, B \, \delta(H - E') \, \eu{-a^2(E-E')^2} \rmd E \, \rmd E' \\
	F_a \equiv \dot{H}_1 &= \iint \delta(H - E) \, F \, \delta(H - E') \, \eu{-a^2(E-E')^2} \rmd E \, \rmd E',
	\label{H1dot}
\end{align}
\end{subequations}
where the second line follows rigorously by virtue that $H$ commutes with the delta functions,
and we have introduced a real-valued parameter $a$ with units of inverse energy.
These expressions can be calculated exactly in the basis of eigenstates of $H$,
which we call $\ket{i}$ with energy $E_i\equiv\braket{i|H|i}$,
and obviously reduce to the traditional approach outlined in the previous section if $a=0$.

We will explain our reasons for choosing this particular form
with an analysis of the behaviour of the modified correlation functions based on \eqn{H1}
in the strict golden-rule limit where a closed-form solution can be derived.
Our findings should also give a good description of the behaviour
in the weak-coupling regime.

It can easily be shown in the eigenstate basis that the important relation $\braket{H_1} = \braket{B}$ holds for any value of $a$.
Therefore, in the slow reaction limit, neglecting $C_{H_1A}(t)$ as before, \cite{Craig2007condensed}
we can compute the rate in terms of the modified flux-side or flux-flux correlation functions
\begin{equation}
	k = Z_A^{-1} C_{F_a B}(t) = Z_A^{-1} \int_0^t C_{F_a F}(\tau) \, \rmd \tau.
	\label{modifiedk}
\end{equation}
This is analogous to the familiar flux correlation function formalism \cite{Miller1983rate}
and thanks to a derivation from the general rate expression \eqn{generalk},
also rigorously gives the correct result for a slow reaction in the long-time limit for any value of $a$ and $\Delta$.
In fact, it can be shown that the long-time limit of the rate formula \eqn{modifiedk}
is unaffected by our introduction of the parameter $a$
by writing the modified flux-flux correlation function 
in the basis of eigenstates:
\begin{align}
	C^{\beta'}_{F_aF}(t) = \sum_{ij} \eu{-\beta E_j + \beta'\Delta E_{ij}} |\!\braket{i|F|j}\!|^2 \, \eu{\iu\Delta E_{ij}t/\hbar} \, \eu{-a^2\Delta E_{ij}^2},
\end{align}
where $\Delta E_{ij}=E_j-E_i$.
Integration over time
using the Fourier relation $\int_{-\infty}^\infty \eu{\iu E t} \rmd t = 2\pi\delta(E)$
shows that the long-time limits of $C^{\beta'}_{F_aB}(t)$ and $C^{\beta'}_{FB}(t)$ are rigorously equivalent.
This therefore proves the equivalence of our modified flux correlation formalism
and the traditional approach derived directly from scattering theory \cite{Miller1983rate}
without the need to use any of the assumptions inherent in linear response theory and nonequilibrium statistical mechanics.

We wish to analyse the effect that our modification has made to the correlation functions at short times,
in particular in the weak-coupling limit.
A simple expression for the modified flux operator, \eqn{H1dot}, in terms of the reactant and product states
can be derived only in the $\Delta\rightarrow0$ limit,
where the eigenstates of $H$ are only slightly perturbed from those of $H_0$,
i.e.\ $\ket{i} \approx \ket{\mu_i}$ or $\ket{\nu_i}$ where $E_i\approx E_{\mu_i}$ or $E_{\nu_i}$.
Any deviation from this approximation leads to higher orders of $\Delta$ and can thus be ignored in the golden-rule limit.
This implies %
\begin{align}
	\delta(H-E)\ket{\mu} &= \sum_i \delta(H-E)\ket{i}\braket{i|\mu} \approx \delta(E_\mu-E)\ket{\mu} %
\end{align}
and its equivalent for product states, $\ket{\nu}$,
such that
\begin{align}
	F_a
	&\approx \frac{\iu}{\hbar} \sum_{\mu\nu} \big(\ket{\mu}\Delta_{\mu\nu}\bra{\nu} - \ket{\nu}\Delta_{\nu\mu}\bra{\mu}\big) \, \eu{-a^2\Delta E_{\mu\nu}^2}.
\end{align}
It is noted that this approximation and hence the remainder of formulae in this section
are only valid in the golden-rule limit and should not be used in general.
The exact expression, \eqn{H1dot}, is used in all the numerical calculations
and the approximation only for the mathematical analysis.

It can now be more clearly seen why the particular form of $H_1$ introduced in \eqn{H1} was chosen.
The parameter $a>0$ can be chosen to ensure that only
the reactant to product transitions which approximately conserve energy
contribute to the modified flux operator.
Note that in the $a\rightarrow\infty$ limit, the exponential becomes sharply peaked like a delta function
such that only strictly energy-conserving transitions are allowed.
It will become apparent that using this limit would have lead directly to the golden-rule rate as described in \secref{TST}
but that the parametrized version given here is more useful,
because from it we can also derive a practical correlation function formalism for dynamical simulations.
We now analyse the correlation functions when using the modified flux.
Describing the dynamics
using first-order perturbation theory, or equivalently by retaining only terms of $O(\Delta^2)$,
which is exact in the golden-rule limit,
we obtain
\begin{align}
	C^{\beta'}_{F_aB}(t)
	&= \frac{\iu}{\hbar} \sum_{\mu\nu} \Tr \Big[ \eu{-\beta' H} \big(\ket{\mu}\Delta_{\mu\nu}\bra{\nu} - \ket{\nu}\Delta_{\nu\mu}\bra{\mu}\big) %
	\nonumber\\ &\quad \times
	\eu{-(\beta-\beta')H} \, \eu{\iu Ht/\hbar} \ketbra{\nu}{\nu} \eu{-\iu Ht/\hbar} \Big] \eu{-a^2\Delta E_{\mu\nu}^2} \\
	&= \frac{1}{\hbar^2} \sum_{\mu\nu} \eu{-\beta E_\mu} |\Delta_{\mu\nu}|^2 \int_0^t \rmd\tau \,  \eu{-a^2\Delta E_{\mu\nu}^2}
	\nonumber\\ &\quad
	\Big[ \eu{-(\beta-\beta')\Delta E_{\mu\nu}}\eu{\iu\Delta E_{\mu\nu}\tau/\hbar}
	+ \eu{-\beta'\Delta E_{\mu\nu}}\eu{-\iu\Delta E_{\mu\nu}\tau/\hbar} \Big].
\end{align}
The modified Kubo-transform is defined from this using \eqn{Ckubo}.

We then assume that the product energy levels are continuous
and that both the density of states and $\Delta_{\mu\nu}^2$ are slowly varying with $E_\nu$
for all transitions allowed from $\ket{\mu}$.
Note that, as for the standard derivation of the golden-rule rate,
by taking the long-time limit, in which the time-integral tends to a sharp peak about $\Delta E_{\mu\nu}=0$,
this assumption can always be made true.
In our case, this approximation is exact (in the golden-rule limit)
not only for $t\rightarrow\infty$
but also at all times in the $a\rightarrow\infty$ limit
when the integrand is localized about $\Delta E_{\mu\nu}=0$ by the Gaussian term.
We therefore write
\begin{align}
	C^{\beta'}_{F_aB}(t)
	&\approx \frac{1}{\hbar^2} \sum_{\mu} \eu{-\beta E_\mu} |\Delta_{\mu\bar{\nu}}|^2 \iint_0^t \rmd\tau \, \rmd\Delta E_{\mu\nu} \, 
	\eu{-a^2\Delta E_{\mu\nu}^2} 
	\nonumber\\ & \quad
	\Big[ \eu{-(\beta-\beta')\Delta E_{\mu\nu}}\eu{\iu\Delta E_{\mu\nu}\tau/\hbar}
	+ \eu{-\beta'\Delta E_{\mu\nu}}\eu{-\iu\Delta E_{\mu\nu}\tau/\hbar} \Big] 
	\\
	&= \frac{1}{\hbar^2} \sum_{\mu} \eu{-\beta E_\mu} |\Delta_{\mu\bar{\nu}}|^2 \frac{\sqrt{\pi}}{a} \int_0^t \rmd\tau
	\nonumber\\ & \quad
	\Big[ \eu{-(\tau + \iu\hbar(\beta-\beta'))^2/4\hbar^2a^2}
	+ \eu{-(\tau -\iu\hbar\beta')^2/4\hbar^2a^2} \Big].
	\label{approxCFaB}
\end{align}
By evaluating the time integrals over the two half-Gaussians
to give $\sqrt{\pi}\hbar a$,
we recover
the golden-rule rate \eqn{goldenrule} for any value of $a$.
This is of course unsurprising as we have derived the result directly from linear response theory
and taken only approximations valid in the $\Delta\rightarrow0$ limit.

This analysis does not however apply at short times with small values of $a$.
In this regime, the approximation that $|\Delta_{\mu\nu}|^2$ and the density of states are slowly varying does not in general hold
and thus
for example the symmetric function in \fig{Kubo} is rather tame
contrary to what would be suggested by \eqn{approxCFaB}.
However, as was shown numerically, the Kubo transform with $a=0$ is nonetheless oscillatory at short times,
as are the symmetric functions in strongly-biased systems. \cite{Huo2013PLDM}
A better indicator of the short-time behaviour is provided by the gradient at $t=0$, equal to $C_{F_aF}(0)$,
which is discussed in Secs.~\ref{sec:dynamics} and \ref{sec:PIGTST}.
From this it is seen that the initial gradient of the flux-side function will decrease as $a$ increases,
which ensures that, for large enough $a$, the function does not overshoot its long-time limit at short times.

It is the presence of imaginary parts in the two exponential terms in \eqn{approxCFaB}
which leads to the oscillatory behaviour at long times observed in \fig{Kubo}.
We note that choosing $\beta'=0$ or $\beta$ will make only one of the two terms non-oscillatory,
and that the symmetrized version, $\beta'=\half\beta$, should give the least problematic form.
The oscillations occur on a time scale of $O(\hbar a^2/\beta)$ with a Gaussian decay on a time scale of $O(\hbar a)$.
The only universal method for damping these oscillations
is to select a large value of $a$ on at least the same order of magnitude as $\beta$
such that the oscillation period is slower than the decay.
In the limit $a\gg\beta$,
\begin{align}
	C_{F_aB}(t)
	&= \frac{1}{\hbar^2} \sum_{\mu} \eu{-\beta E_\mu} |\Delta_{\mu\bar\nu}|^2 \frac{\sqrt{\pi}}{a} 2 \int_0^t \eu{-\tau^2/4\hbar^2a^2} \rmd\tau \\
	&= \frac{2\pi}{\hbar} \sum_{\mu} \eu{-\beta E_\mu} |\Delta_{\mu\bar\nu}|^2 \Erf(t/2\hbar a) .
	\label{CFaB-Erf}
\end{align}
The error function, $\Erf(z)$, tends to 1 in the $z\rightarrow\infty$ limit
and, as expected, recovers the golden-rule rate.
As this result does not depend on the value of $\beta'$,
we have dropped the superscript from the correlation function.

\begin{figure}
	\includegraphics{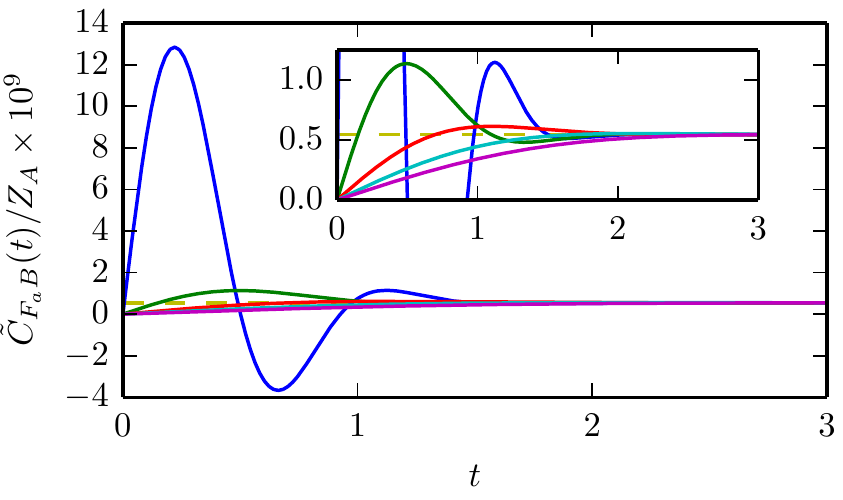}
	\caption{The Kubo-transformed modified flux-side correlation function for various values of $a$:
	blue $a=0.2$, green $a=0.4$, red $a=0.6$, cyan $a=0.8$, magenta $a=1$.
	All five converge to the golden-rule rate (yellow dashed) in the long time limit as can be seen in the magnified inset.
	Note that the $a=0$ version, shown in \fig{Kubo}, is way off this scale.
	}
	\label{fig:unbiased}
\end{figure}

Of course, this means that for very large $a$, it will be necessary to propagate to longer times to reach the plateau.
Therefore for each problem, we should choose the minimum value of $a$ which defines a non-oscillatory function
in order to achieve the best efficiency.
We explore the effects of varying this parameter numerically in \fig{unbiased}.
It is seen that the Kubo correlation function, which had large oscillations for $a=0$
has been smoothed out using the parameter $a>0$,
where it tends to the shape of an error function.

\begin{figure}
	\includegraphics{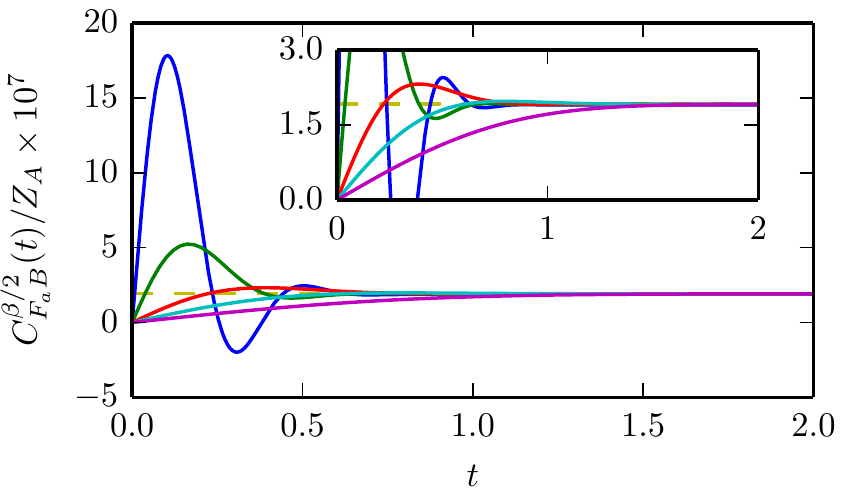}
	\caption{The symmetric correlation functions ($\beta'=\half\beta$) for a biased system with
	$\beta=1$, $\alpha=0.5$, $\epsilon=10$ for various values of $a$:
	blue $a=0$, green $a=0.1$, red $a=0.2$, cyan $a=0.3$, magenta $a=0.5$.
	All five converge to the golden-rule rate (yellow dashed) in the long time limit as can be seen in the magnified inset.
	}
	\label{fig:biased}
\end{figure}

In the biased case, even the symmetric correlation function becomes oscillatory as shown in \fig{biased}.
We note that varying the value of $\beta'$ only makes the problem worse,
but that increasing $a$ removes the oscillations completely.

This completes our definition and study of the non-oscillatory flux-side correlation function formalism.
In summary,
due to the rigorous derivation from linear response theory,
our modifications have not changed the long-time limit of the correlation functions
from which the rate constant is defined,
regardless of the strength of the electronic coupling parameter $\Delta$.
However, the particular form of nonequilibrium perturbation used, \eqn{H1},
has removed the oscillatory problem in the weak-coupling limit.
In the adiabatic, strong-coupling limit,
the most efficient way to compute the rate is of course with none of these functions
but with a position-dependent dividing surface.
A complete nonadiabatic rate theory should be able to make use of both of these limits.

Note that we have not simply computed the oscillatory integral using a stationary-phase approximation
but instead derived a non-oscillatory version which gives the exact rate.
This is not only a more accurate approach
but also leads to useful new developments.
The non-oscillatory correlation function formalism is expected to be of great use in the development of new approximate dynamical methods.
Immediate advances include
an improved initial distribution for nonadiabatic classical trajectory simulations
and a new formulation of the exact golden-rule rate in terms of the quantum trace. %
We deal with the former in the next section and the latter in \secref{TST}.

\section{Modified initial distributions}
\label{sec:dynamics}

As we have already discussed,
a trajectory simulation of the traditional Kubo-transformed flux-side correlation function
in the nonadiabatic, weak-coupling limit
will be dominated by transitions occurring in the reactant or product wells
leading to strong oscillations at short times.
We know from the laws of energy conservation,
that successful transitions should only occur 
in the region of the crossing point.
However, neither the flux, $F$, nor a Kubo-transformed thermal flux operator localizes the initial distribution here.
As we shall show, the modified flux, $F_a$, offers the possibility to alleviate this problem via the strength of the parameter $a$.

We shall consider here the classical limit of trajectory-based nonadiabatic dynamics simulations of one-dimensional systems,
as this provides one of the simplest problems that poses a significant challenge to current methods. \cite{Landry2012hopping,Menzeleev2014kinetic}
By this limit,
as discussed in \Ref{Tully1990hopping},
we imply that the system is formed of heavy nuclei
moving with small velocities
but which are still allowed to hop between diabatic surfaces---an inherently quantum effect.
We do not specify any particular form of trajectory method,
or even if it gives exact or approximate dynamics.
The extension to treat multidimensional systems with position-dependent electronic coupling is trivial
and we leave the generalization for the treatment of quantum nuclei by path-integral methods to future work.

Let us assume that we wish to compute 
the Kubo-transformed modified flux-flux correlation function, $\tilde{C}_{F_aF}(t)$.
The distribution of the initial nuclear coordinate contributing to the short-time limit is given by
\begin{align}
	\tilde{P}_a(x)
	&= \frac{1}{\beta} \int_0^\beta \Tr \left[ \eu{-\beta' H} F_a \, \eu{-(\beta-\beta')H} F \, \delta(\op{x}-x) \right] \rmd\beta',
	\label{Paquantum}
\end{align}
such that $\tilde{C}_{F_aF}(0)=\int_{-\infty}^\infty \tilde{P}_a(x) \, \rmd x$.
In order to put this into closed form, where we can more easily study the properties of the distribution,
we substitute quantum operators for localized classical variables
to give an approximation to the modified flux
\begin{align}
	F_a
	&\approx \frac{\iu\Delta}{\hbar} \iint \delta(p^2/2m + V(x)-E) \, \big( \ketbra{0}{1} - \ketbra{1}{0} \big)
	\nonumber\\ & \quad \times
	\delta(p^2/2m + V(x)-E') \, \eu{-a^2(E-E')^2} \rmd E \, \rmd E'.
\end{align}
The integrals can be performed at each value of $x$ in 
the basis of adiabatic electronic states $\ket{\chi_\pm(x)}$,
which diagonalize the diabatic potential matrix $V(x)$.
The corresponding eigenvalues are $W_\pm(x)\equiv\braket{\chi_\pm(x)|V(x)|\chi_\pm(x)}$.
Using the relations given in Appendix~\ref{sec:adiabatic}
along with standard trigonometric identities,
we can show that it simplifies to give
\begin{equation}
	F_a \approx \frac{\iu\Delta}{\hbar} \big( \ketbra{0}{1} - \ketbra{1}{0} \big) \, \eu{-a^2[W_+(x)-W_-(x)]^2}.
\end{equation}
It is seen that the modified flux, within the classical approximation,
is simply the standard flux operator weighted by a Gaussian which localizes it about the avoided crossing.

The classical distribution contributing to $\tilde{C}_{F_aF}(0)$
is obtained in the same way %
to give
\begin{multline}
	\tilde{P}_a(x) 
	\approx \sqrt\frac{m}{2\pi\beta\hbar^2} \frac{2\Delta^2}{\beta\hbar^2}
		\frac{\eu{-\beta W_-(x)} - \eu{-\beta W_+(x)}}{W_+(x) - W_-(x)}
		\\\times
		\eu{-a^2[W_+(x)-W_-(x)]^2}.
	\label{Pacl}
\end{multline}
In the golden-rule limit, i.e.\ retaining only terms of order $\Delta^2$, this becomes
\begin{multline}
	\tilde{P}_a(x) 
	\approx \sqrt\frac{m}{2\pi\beta\hbar^2} \frac{2\Delta^2}{\beta\hbar^2}
		\frac{\eu{-\beta V_1(x)} - \eu{-\beta V_0(x)}}{V_0(x) - V_1(x)}
		\\\times
		\eu{-a^2[V_0(x)-V_1(x)]^2}.
		\label{PaGR}
\end{multline}

\begin{figure}
	\includegraphics{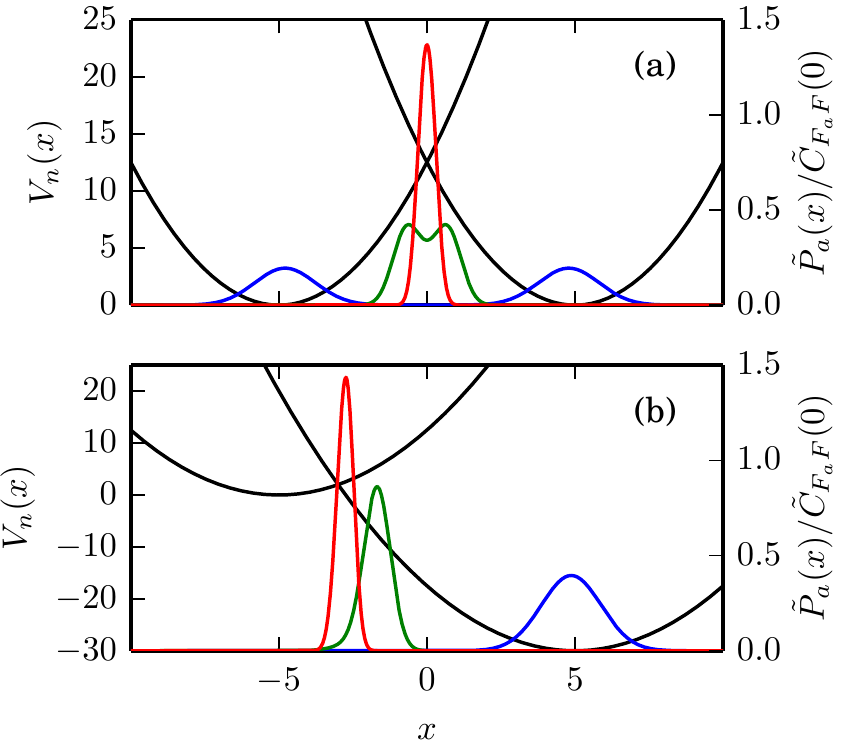}
	\caption{Normalized classical nuclear distributions, $\tilde{P}_a(x)$, in the golden-rule limit, \eqn{PaGR},
	for initializing a Kubo-transformed modified flux-flux correlation function calculation;
	$a$ is equal to 0 (blue), 0.15 (green) or 0.3 (red).
	The system is a typical two-state curve-crossing problem, \eqn{Marcus}, with diabatic surfaces shown in black, $\beta=1$
	no bias in the top panel and $\epsilon=30$ in the lower.
	}
	\label{fig:distribution}
\end{figure}

It is from distribution functions such as these that initial points are selected for trajectory simulations.
\cite{*[{The sampling of a correlation function using a distribution based on its zero-time limit is discussed in }] [{}] Zimmermann2013sampling}
Note that in order to sample these functions,
it would not be necessary to know the exact location of the crossing of the potentials
which for a complex multidimensional system would be a high-dimensional hypersurface
and is generally difficult to calculate.
Instead, a Monte Carlo evaluation would automatically select samples from the correct region of configuration space.
The distribution at the same level of approximation
corresponding to the original unmodified Kubo-transformed flux-flux correlation function, $\tilde{P}_0(x)$, is found by setting $a=0$.
A range of approximate modified distributions with $a\ge0$ are presented in \fig{distribution} for a biased and an unbiased system.

It is seen that large values of $a$ localize the nuclear distribution about the crossing point
with an effective localization given by
\begin{equation}
	\frac{\tilde{P}_a(x)}{\tilde{P}_0(x)} \approx \eu{-a^2[W_+(x)-W_-(x)]^2}.
\end{equation}
This effect applies equally to symmetric as to biased systems even deep into the inverted regime.
This is exactly the effect that we wished to apply to the system
based on an intuition that trajectories should hop between diabatic surfaces only in this region.
Estimates for good values of $a$ can be obtained from
an analysis of the golden-rule limit of the approximate distribution function, given by \eqn{PaGR}.
We here assume that the system has strong vibronic coupling such that $\beta V_n'(x^\ddagger)^2 \gg V_n''(x^\ddagger)$,
where $x^\ddagger$ is the crossing point such that $V_0(x^\ddagger)=V_1(x^\ddagger)$.
The distribution in a symmetric system has a maximum at $x=0$ for $a\ge\beta/\sqrt{24}$.
For an asymmetric system, the distribution at $x^\ddagger$ is not necessarily a stationary point
but has a negative curvature,
and is therefore well within the envelope,
for $a\ge\beta\sqrt{1/6+V_0'(x^\ddagger)V_1'(x^\ddagger)/2\mathcal{F}^2}$,
where $\mathcal{F}=|V_0'(x^\ddagger)-V_1'(x^\ddagger)|$.
We note that the value of $a$ used in simulations can be chosen to minimize oscillations and optimize the efficiency of the calculation
and that these estimates may be used for a good first guess and do not need to be accurately calculated.
In principle, the rate constant found by an accurate simulation will not depend on the value chosen.

With an exact dynamical method, one could therefore compute the modified flux-flux correlation function,
and hence the rate constant,
starting from a localized region of configuration space.
Note that the localization has appeared naturally from our derivation from linear response theory.
We could not have simply performed Monte Carlo importance sampling of $\tilde{P}_0(x)$ with an umbrella function centred about the crossing point
because this would still be a simulation of the original correlation function,
and in fact the effects of outlying trajectories which do start in the wells would be magnified by the umbrella and worsen the statistics even more.
Although a nuclear-configurational dividing surface could also be used to localize the nuclei at the crossing point,
as discussed above, such dividing-surface approaches lead to inefficient calculations
due to its time derivative, the flux, which depends on the nuclear momenta rather than diabatic hops.

It was possible for previous studies \cite{Topaler1996nonadiabatic,Wang1999mapping,Wang2006flux}
to perform efficient simulations of \textit{symmetric} correlation functions
without the present modifications.
This is because the symmetrized thermal flux $\eu{-\beta H/2}F\eu{-\beta H/2}$, unlike its Kubo-transformed version,
is localized about $x=0$ for a spin-boson model.
However, for strongly biased systems, \cite{Menzeleev2011ET,Huo2013PLDM}
this does not coincide with the diabatic crossing point, $x^\ddagger$,
and the current theory may be useful to avoid strongly oscillatory functions.

In conclusion, the proposed modified flux-side correlation functions $C_{F_aB}(t)$
can be used instead of the original flux-side versions $C_{FB}(t)$
as they all tend to the same long-time limit from which the rate constant is given exactly.
Therefore, for an exact dynamical method %
based on flux correlation functions, and in particular the Kubo-transformed versions,
the proposed modifications will improve the efficiency without affecting the result,
regardless of the strength of electronic coupling.
Approximate methods will also see a large efficiency gain
and may even see an improvement in the accuracy of their results
as phase cancellation is no longer necessary with this more intuitive localized initial distribution.
Such applications will be explored in future work.

\section{Golden-rule transition-state theory}
\label{sec:TST}
\subsection{Quantum formulation}
\label{sec:PIGTST}

The modified nonequilibrium perturbation, \eqn{H1}, was introduced in order to damp the oscillations of the flux correlation functions,
but we shall show that it can also be used
to derive a new formulation of a nonadiabatic quantum transition-state theory, which is exact in the golden-rule limit.

There are many different theories in the literature bearing the TST label,
often with quite different meanings.
We use the definition that TST is
a dynamics-free approach based on the statistical mechanics of a
low-probability region of phase space associated with transitions---in this case, near the crossing point where the diabatic potentials are equal.
There must also exist a related correlation function from which the exact rate can be computed
and which is equivalent to the TST rate if there is no recrossing in the dynamics.

The definition of recrossing in the correlation function varies depending on the system studied.
For example, for adiabatic dynamics, a dividing surface in nuclear-configuration space is usually defined
and the TST assumption is that trajectories will never cross this dividing surface more than once.
For the classical Born-Oppenheimer flux-side correlation function \cite{Chandler1978TST,ChandlerGreen}
and a particular generalization of its quantum equivalent, \cite{Hele2013QTST,*Althorpe2013QTST}
it can be shown that the non-recrossing assumption leads to a step-like shape
and therefore that the TST rate is proportional to its $t\rightarrow0_+$ limit.

In our case, where there is no dividing surface,
a recrossing path will be defined as one which hops between the diabatic states more than once
within either the forward or backward propagators $\eu{\mp\iu H t/\hbar}$.
Because each hop introduces a factor of $\Delta$, this non-recrossing assumption is exact in the golden-rule limit. 
For a system with a continuum of product states in the $\Delta\rightarrow0$ limit, %
it was seen in \eqn{CFaB-Erf} that
the correlation function $C_{F_aB}(t)$ with $a\rightarrow\infty$ goes like the error function
and that therefore its time-derivative,
the modified flux-flux correlation function,
$C_{F_aF}(t)$ is Gaussian with known width.
We can therefore use this as our correlation function from which the exact rate can be computed whether
the non-recrossing assumption is valid or not,
but which, in the golden-rule limit, has a simple form
from which we can compute the TST rate solely from its value at $t=0$.
As was shown in the previous section, the corresponding distribution will be localized about the crossing region.
Like other quantum TSTs,\cite{Hele2013QTST} our formulation will not necessarily be variational,
i.e.\ the exact rate may be smaller or larger than the TST rate.

According to our definition,
the approach of Wolynes \cite{Wolynes1987nonadiabatic} %
is not a TST but instead a quantum instanton approximation
as it is based on a \textit{steepest-descent} integration along time. \cite{Vanicek2005QI}
It is therefore not necessarily exact for a system with no recrossing, that is in the golden-rule $\Delta\rightarrow0$ limit,
and is not directly related to a dynamical method.

In order to formulate a quantum expression for the golden-rule TST rate,
we utilize the fact that we know the shape of the modified flux-flux correlation function in the limit $a\rightarrow\infty$
within the non-recrossing assumption.
That is
\begin{align}
	\lim_{a\rightarrow\infty} C_{F_aF}(t) &\approx \lim_{a\rightarrow\infty} C_{F_aF}(0) \, \eu{-t^2/4\hbar^2a^2},
\end{align}
where the approximation becomes exact in the $\Delta\rightarrow0$ limit.
The value at zero time
for any value of $\beta'$
is given by
\begin{align}
	C_{F_aF}(0)
	&= \Tr \left[ \eu{-\beta'H} F_a \, \eu{-(\beta-\beta')H} F \right] \\
	&\equiv \frac{2\sqrt{\pi}\Delta^2}{\hbar^2 a} \mathcal{Z}^\ddagger(a)
\end{align}
where $\mathcal{Z}^\ddagger(a)$ is defined by this equivalence.
We introduce the transition-state partition function
\begin{align}
	Z^\ddagger
	&= \lim_{a\rightarrow\infty} \mathcal{Z}^\ddagger(a) \\
	&= \lim_{a\rightarrow\infty} \frac{a}{2\sqrt{\pi}} \iint \rmd E \, \rmd E' \, \eu{-a^2(E-E')^2} 
		\lim_{\Delta\rightarrow0}
		\nonumber\\ &\quad
		\Tr \Big[
		\eu{-\beta'H} \delta(H-E) \ketbra{0}{1}
		\delta(H-E') \, \eu{-(\beta-\beta')H} \ketbra{1}{0}
		\nonumber\\ &\quad
		+ \eu{-\beta'H} \delta(H-E) \ketbra{1}{0}
		\delta(H-E') \, \eu{-(\beta-\beta')H} \ketbra{0}{1} \Big]
		\\
	&= \int \eu{-\beta E} \lim_{\Delta\rightarrow0} \Tr \big[ \delta(H-E) \ketbra{0}{1} \delta(H-E) \ketbra{1}{0} \big] \rmd E \\
	&= \int \eu{-\beta E} \tr \big[ \delta\big(\!\braket{0|H|0}-E\big) \, \delta\big(\!\braket{1|H|1}-E\big) \big] \rmd E,
	\label{Zddagger}
\end{align}
in which we have used a relation for the Dirac delta function
$\delta(z)=\lim_{a\rightarrow\infty} \frac{a}{\sqrt{\pi}} \eu{-a^2z^2}$
and performed the integration over $E'$.
In the last line, the quantum trace is taken only over nuclear degrees of freedom.

The golden-rule TST rate is therefore
\begin{align}
	k_\mathrm{TST}
	&= Z_A^{-1} \lim_{a\rightarrow\infty} \int_0^\infty C_{F_aF}(0) \, \eu{-\tau^2/4\hbar^2a^2} \rmd \tau \\
	&= \frac{2\pi\Delta^2}{\hbar} \frac{Z^\ddagger}{Z_A}.
	\label{kTST}
\end{align}
This expression is only valid in the $\Delta\rightarrow0$ limit
where it gives the quantum golden-rule result exactly
but without explicitly using the states of the system.
This is most easily seen by evaluating the trace in the basis of the vibronic states before integrating over energy
from which \eqn{goldenrule} is recovered.
It is however a more general result 
which also applies to complex systems
after a trivial extension to multidimensional nuclear configurations
and non-constant electronic coupling $\Delta(\op{x})$,
which would then be included inside the trace of \eqn{Zddagger}.

We could have derived this result
from an alternative formulation of the rate constant
in terms of the microcanonical cumulative reaction probability,
equivalent to that of \Ref{Miller1983rate} with a substitution for the electronic flux, \eqn{flux}:
\begin{equation}
	k = \frac{\pi\hbar}{Z_A} \int \eu{-\beta E} \Tr \big[F\delta(E-H)F\delta(E-H)\big] \rmd E.
\end{equation}
This general formulation gives the exact rate constant regardless of the coupling strength, $\Delta$,
and reduces to our TST result, \eqn{kTST}, in the $\Delta\rightarrow0$ limit.
However, the new derivation presented in this work has shown that the golden-rule result can be considered a transition-state theory
in the sense that it can be linked to the short-time limit of the modified flux-flux correlation function
in the $a\rightarrow\infty$ limit.
Such links with transition-state theories can be invaluable when developing new trajectory-based methods for rate calculations.
\footnote{An important example is RPMD and its TST limit described in
\Refs{RPMDrefinedRate,rpinst,Hele2013QTST,Althorpe2013QTST}
}

The new formulation, \eqn{kTST}, can be compared to the imaginary-time path-integral methods for computing the golden-rule rate
of
\Refs{Wolynes1987nonadiabatic,Cao1997nonadiabatic}.
They also depend on a transition-state partition function where two diabatic hops are enforced
but without explicitly imposing energy conservation.
These methods were derived from a steepest-descent approximation to the analytically-continued flux-flux correlation function formalism
and are not exact, even in the golden-rule limit.
Our new formulation does not take these approximations and also avoids the unsavoury analytic continuation.

We note that unlike the former approaches,
for which results are efficiently computed using path-integral Monte Carlo,
at first glance, a numerical evaluation of \eqn{Zddagger} using path integrals
\cite{Chandler+Wolynes1981,Wolynes1987nonadiabatic,Alexander2001diabatic}
or semiclassical instanton methods \cite{Miller1975semiclassical,Cao1997nonadiabatic}
looks extremely complicated due to presence of the microcanonical density operators.
\cite{Miller1975rate,Lawson2000microcanonical}
However, it is simplified somewhat by the integral over energy.
Further work will assess whether a practical formulation can be found applicable to complex systems.
We can however analyse quite easily the classical limit which is outlined in the next section.

\subsection{Classical golden-rule rate}
\label{sec:classical}

In the limiting case that the one-dimensional nuclear motion can be considered classically,
we can show that the TST formulation derived above
gives the same rate constant as Landau-Zener theory
or, in the special case of a spin-boson system, Marcus theory.

We take the classical limit of \eqn{Zddagger} %
as in \secref{dynamics}
to give
\begin{align}
	Z^\ddagger
	\label{Zddaggercl1}
	&= \frac{1}{2\pi\hbar} \int \rmd E \, \eu{-\beta E}
	\iint_{-\infty}^\infty \rmd x \, \rmd p
	\nonumber\\ &\quad \times
	\delta\big(p^2/2m + V_0(x) - E\big) \, \delta\big(p^2/2m + V_1(x) - E\big)  \\
	\label{Zddaggercl}
	&= \frac{1}{2\pi\hbar} \iint_{-\infty}^\infty \eu{-\beta p^2/2m} \, \eu{-\beta V_0(x)} \, \delta\big[V_0(x) - V_1(x)\big] \, \rmd x \, \rmd p,
\end{align}
where here $x$ and $p$ are considered as classical variables.
This same result could also be derived from the classical limit \cite{Lawson2000microcanonical}
of the elements of the quantum microcanonical density matrix. \cite{Miller1975rate}
Note that the two kinetic energy terms are equal in \eqn{Zddaggercl1},
and have therefore cancelled out in the delta function in \eqn{Zddaggercl}.
This is because the flux operator does not affect position or momentum, only electronic states.

We proceed using
\begin{align}
	\delta\big[V_0(x) - V_1(x)\big] = \frac{\delta(x - x^\ddagger)}{|V_0'(x^\ddagger) - V_1'(x^\ddagger)|}
\end{align}
and by defining the activation energy as 
$V^\ddagger \equiv V_0(x^\ddagger) = V_1(x^\ddagger)$
to give, from \eqn{kTST}, a general formula for the classical TST rate,
\begin{align}
	k_\mathrm{cl,TST}
	&= \sqrt{\frac{2\pi m}{\beta\hbar^2}} \frac{\Delta^2}{\hbar|V_0'(x^\ddagger)-V_1'(x^\ddagger)|} \frac{\eu{-\beta V^\ddagger}}{Z_A},
	\label{clTST}
\end{align}
which is the same result as found by the Landau-Zener approach in the golden-rule limit for a one-dimensional system \cite{Nitzan}
\begin{align}
	k_\mathrm{LZ}
	&= \int_{-\infty}^\infty |\dot{x}| P(x^\ddagger,\dot{x}) P_{0\rightarrow 1}^\ddagger(\dot{x}) \, \rmd\dot{x},
\end{align}
with \cite{Zener1932LZ} %
\begin{align}
	P(x^\ddagger,\dot{x}) &= \frac{m}{2\pi\hbar Z_A} \, \eu{-\frac{1}{2} \beta m \dot{x}^2} \eu{-\beta V^\ddagger} \\
	P^\ddagger_{0\rightarrow 1}(\dot{x}) &= \frac{2\pi\Delta^2}{\hbar|\dot{x}||V'_0(x^\ddagger) - V_1'(x^\ddagger)|}.
\end{align}
Note that we have assumed that reactive trajectories can occur with positive or negative velocity, $\dot{x}$,
as is appropriate for bound reactant states.
\footnote{In fact this also holds for scattering states bound on the product side
such that classical trajectories must change direction and come back via the crossing point $x=x^\ddagger$.}

As a specific example, we consider a model system typically used in Marcus' theory of electron-transfer reactions,
defined as in \eqn{modelH} but with two harmonic potential energy surfaces \cite{Weiss}
\begin{subequations}
\label{Marcus}
\begin{align}
	V_{0}(x) &= \half m \omega^2 (x + x_0)^2, \\
	V_{1}(x) &= \half m \omega^2 (x - x_0)^2 - \epsilon.
\end{align}
\end{subequations}
The reorganization energy is $\Lambda=2m\omega^2x_0^2$
and the crossing point is
$x^\ddagger=-x_0\epsilon/\Lambda$
with
$V^\ddagger = (\Lambda-\epsilon)^2/4\Lambda$
and
$V_0'(x^\ddagger) - V_1'(x^\ddagger) = \Lambda/x_0$.
The reactant partition function is $Z_A=(\beta\hbar\omega)^{-1}$ %
and the rate constant predicted by \eqn{Zddaggercl} is
\begin{align}
	k_\mathrm{cl,TST}
	&= \frac{\Delta^2}{\hbar} \sqrt{\frac{\pi \beta}{\Lambda}} \, \eu{-\beta (\Lambda - \epsilon)^2 / 4\Lambda}.
\end{align}
This is equal to the familiar classical Marcus rate, \cite{Marcus1956ET}
which is the result for a spin-boson system in the golden-rule limit with classical nuclei.

This link to Marcus theory is perhaps more exciting than it at first seems
because, as already discussed,
the path-integral golden-rule rate formula proposed by Wolynes, \cite{Wolynes1987nonadiabatic} Cao and Voth \cite{Cao1997nonadiabatic}
cannot describe the correct behaviour in the inverted regime $\epsilon>\Lambda$.
This shows that the current formulation, \eqn{kTST}, is a more powerful 
starting point for the derivation of nonadiabatic golden-rule TST
than the analytic-continuation or \ImF\ methods.

Although the classical formula has reduced to well-known results,
its formulation offers something new,
which is a direct link between the non-dynamical Marcus theory
and an exact correlation function formalism,
showing that Marcus theory itself, and its anharmonic generalization \eqn{clTST},
can be thought of as classical nonadiabatic transition-state theories.
This implies that if a classical trajectory-based dynamical method, which may be efficiently initialized near the crossing-point,
gives the correct Gaussian decay of the modified flux-flux correlation function in the large $a$ limit,
it will necessarily reproduce rates of Marcus theory when in the golden-rule limit.
This sets a very clear goal for the development of new nonadiabatic dynamical theories.

We note also that a multidimensional extension to \eqn{Zddaggercl} is found trivially by following the same procedure,
which when combined with the usual efficient approaches for the calculation of free energies,
gives perhaps one of the simplest methods for computing classical golden-rule rates
without making the harmonic approximation.

\section{Conclusions}
\label{sec:conclusions}

We have analysed the use of the Kubo-transformed flux-side correlation function,
with reactants and products defined as usual by projections onto electronic states, \cite{Topaler1996nonadiabatic}
to compute the nonadiabatic rate constant for systems with weak electronic coupling.
It was shown to be very inefficient due to strong oscillatory behaviour
stemming from spurious diabatic transitions occurring between low- and high-energy states,
which due to phase cancellation at long times do not contribute to the rate constant.
Using a formalism based on linear response theory,
we have proposed a modified flux-side correlation function
which rigorously gives the same exact rate in the long-time limit,
regardless of the electronic coupling strength,
but which includes a parameter $a$ which can be chosen to 
remove the oscillations at short times.
An alternative derivation equates the formalism directly with quantum scattering theory
via the traditional flux correlation functions. \cite{Miller1983rate}

Non-zero values of the parameter change the nuclear distribution function used to initialize typical trajectory simulations
such that the distribution is localized about the crossing of the potential energy surfaces.
Very large values of $a$ may need a longer propagation time to reach a plateau
so we recommend a medium value of $a\sim\beta$
which would give an initial distribution of nuclear configurations distributed close to the crossing
where the electronic states have similar energies.
For complex systems, it will not of course be possible to compute the modified flux correlation functions exactly,
any more than it is for the original form.
However, the new formalism provides a rigorous base on which to build approximate dynamical methods,
which share the benefits of being non-oscillatory and of starting from a distribution localized about the crossing point.
Such methods will be more efficient and perhaps even more accurate due to the proposed modifications.

The new formalism is of particular importance for the use of nonadiabatic RPMD \cite{mapping} in the weak-coupling regime,
which like all ring-polymer approaches, \cite{RPMDcorrelation}
approximately computes Kubo-transformed correlation functions.
The Kubo transform is also the most appropriate form for other classical-trajectory approaches
as it shares many symmetry properties with classical correlation functions. \cite{RPMDcorrelation}
Further work will attempt to compute the non-oscillatory flux-side correlation functions using such approaches.
However, it may also be possible to use
the proposed function to improve the efficiency of a wide range of other nonadiabatic dynamical methods
\cite{Wang2006flux,Topaler1996nonadiabatic,Wang1999mapping}
based on the symmetric correlation function %
at least when studying strongly biased systems.
\cite{Menzeleev2011ET,Huo2013PLDM}

Our study of the initial distribution for trajectory simulations
was based on a classical limit.
Path-integral methods could be used for example to initialize nonadiabatic ring-polymer dynamics
from \eqn{Paquantum}
but further work is required before a practical path-integral sampling scheme can be implemented.
We note, however, that even without this extension,
there is still a large applicability for trajectory methods for the nonadiabatic dynamics of classical nuclei,
using for example surface-hopping or classical mapping approaches.
In fact, even the nonadiabatic RPMD method \cite{mapping} with classical nuclei
is worthy of study because 
the introduction of ring-polymer beads was seen not only to describe nuclear quantum effects but also to improve the electronic dynamics;
i.e.\ it does not reduce to the classical $N=1$ mapping approach when the ring polymers collapse in nuclear-configuration space.
There is therefore good reason to attempt to utilize such methods to compute the modified flux-side correlation functions proposed here.

We were also able to extract another useful development 
from the non-oscillatory correlation function formalism,
which is the derivation of an expression for a nonadiabatic TST rate constant,
exact in the golden-rule, $\Delta\rightarrow0$, limit.
This limit is often encountered in electron-transfer processes in the condensed phase, \cite{ChandlerET}
the simulation of which would provide many possible applications for the method.
If the nuclei are considered to be classical and are treated in the harmonic approximation,
the expression reduces to the Marcus theory rate even in the inverted regime.
Further work is needed to find a practical expression for computer simulation
which will allow the efficient computation of nonadiabatic rates in the golden-rule limit for complex systems
including quantum-mechanical effects such as nuclear tunnelling in proton-coupled electron transfer. \cite{HammesSchiffer2010PCET}

\section{Acknowledgement}
The authors would like to thank Stuart C. Althorpe for helpful comments on the manuscript.
JOR gratefully acknowledges a Research Fellowship from the Alexander von Humboldt Foundation.

\appendix

\section{Vibronic states}
\label{sec:vibronic}

We define the one- or multidimensional nuclear configuration as $x$
and the positions of all electrons in the system as $\mathbf{r}$.
As these variables cover all spatial degrees of freedom in a molecular system, we recognize the identity
$\iint \ketbra{x,\mathbf{r}}{x,\mathbf{r}} \rmd x \, \rmd\mathbf{r} = 1$.

Writing the electronic (diabatic) wave functions as
$\phi_0(\mathbf{r};x)$ and $\phi_1(\mathbf{r};x)$
and nuclear wave functions as
$\psi_\mu(x)$ and $\psi_\nu(x)$,
we can define the vibronic states using
\begin{align*}
	\braket{x,\mathbf{r}|\mu} &= \psi_\mu(x) \phi_0(\mathbf{r};x), &
	\braket{x,\mathbf{r}|0} &= \bra{x} \phi_0(\mathbf{r};x), \\
	\braket{x,\mathbf{r}|\nu} &= \psi_\nu(x) \phi_1(\mathbf{r};x), &
	\braket{x,\mathbf{r}|1} &= \bra{x} \phi_1(\mathbf{r};x).
\end{align*}

Projections onto the electronic states give
\begin{align*}
	\braket{0|\mu}
	&= \int \ket{x} \psi_\mu(x) \, \rmd x, &
	\braket{1|\nu}
	&= \int \ket{x} \psi_\nu(x) \, \rmd x.
\end{align*}
The coupling matrix element can therefore be defined as
\begin{align}
	\braket{\mu|0} \braket{1|\nu}
	&= \iint \psi_\mu(x')^* \braket{x'|x} \psi_\nu(x) \, \rmd x \, \rmd x' \\
	&= \int \psi_\mu(x)^* \psi_\nu(x) \, \rmd x,
	\label{couplingintegral}
\end{align}
which is the Franck-Condon matrix element.

\section{Wave functions of the model system}
\label{sec:wavefunctions}

The reactant well, \eqn{Harmonic}, has the form of a harmonic oscillator
with eigenstates
\begin{align}
	\psi_{\mu}(x)
	= \left(\frac{m\omega}{\pi\hbar}\right)^\frac{1}{4} \frac{1}{\sqrt{2^\mu \mu!}} \, \eu{-m\omega x^2/2\hbar} H_\mu\left(\sqrt{\frac{m\omega}{\hbar}}x\right),
\end{align}
where $\mu\in\{0,1,\dots\}$
and $H_\mu$ are the Hermite polynomials.
The discrete energies are $E_\mu = (\mu+\half)\hbar\omega$.

The product wave functions are the continuum states of the repulsive Morse oscillator, \eqn{Morse},
normalized in energy space, \cite{Muendel1984morse}
\begin{align}
	\psi_{\nu}(x)
	= \sqrt{\frac{m\hbar}{2\pi p y}} (2b)^{-\zeta} \frac{\Gamma(\half-\zeta)}{\Gamma(-2\zeta)} W_{0,\zeta}(y),
\end{align}
for $E_\nu > -\epsilon$,
where \mbox{$p=\sqrt{2m(E_\nu+\epsilon)}$},
\mbox{$b=\sqrt{2mD}/\hbar\alpha$},
\mbox{$\zeta=\iu p/\hbar\alpha$}
and
\mbox{$y=2b\exp[-\alpha(x-x_0)]$}.
This particular Whittaker function is related \cite{[{See Eq.\,9.235.2 from }] [{}] Gradshteyn}
to a modified Bessel function of purely imaginary order,
\mbox{$W_{0,\zeta}(y)=\sqrt{y/\pi}K_\zeta(y/2)$},
which can be can be computed with the algorithm of \Ref{Gil2004iBessel}.

\section{Adiabatic states}
\label{sec:adiabatic}

The adiabatic states $\ket{\chi_\pm(x)}$ are defined in the diabatic basis
as the eigenvectors of the $2\times2$ diabatic matrix $V(x)$, given by \eqn{Vmatrix}.
One can show \cite{Tannor} that
\begin{align*}
	\braket{0|\chi_+(x)} &= \cos\tfrac{\theta}{2} & \braket{0|\chi_-(x)} &=-\sin\tfrac{\theta}{2} \\
	\braket{1|\chi_+(x)} &= \sin\tfrac{\theta}{2} & \braket{1|\chi_-(x)} &= \cos\tfrac{\theta}{2} 
\end{align*}
is the solution with
\begin{align}
	\tan\theta = \frac{2\Delta}{V_0(x) - V_1(x)}.
\end{align}
The corresponding eigenvalues are
\begin{align}
	W_\pm(x) = \frac{V_0(x) + V_1(x)}{2} \pm \half \sqrt{[V_0(x)-V_1(x)]^2 + 4\Delta^2}.
\end{align}

\bibliography{../references} %

\begin{thebibliography}{95}%
\makeatletter
\providecommand \@ifxundefined [1]{%
 \@ifx{#1\undefined}
}%
\providecommand \@ifnum [1]{%
 \ifnum #1\expandafter \@firstoftwo
 \else \expandafter \@secondoftwo
 \fi
}%
\providecommand \@ifx [1]{%
 \ifx #1\expandafter \@firstoftwo
 \else \expandafter \@secondoftwo
 \fi
}%
\providecommand \natexlab [1]{#1}%
\providecommand \enquote  [1]{``#1''}%
\providecommand \bibnamefont  [1]{#1}%
\providecommand \bibfnamefont [1]{#1}%
\providecommand \citenamefont [1]{#1}%
\providecommand \href@noop [0]{\@secondoftwo}%
\providecommand \href [0]{\begingroup \@sanitize@url \@href}%
\providecommand \@href[1]{\@@startlink{#1}\@@href}%
\providecommand \@@href[1]{\endgroup#1\@@endlink}%
\providecommand \@sanitize@url [0]{\catcode `\\12\catcode `\$12\catcode
  `\&12\catcode `\#12\catcode `\^12\catcode `\_12\catcode `\%12\relax}%
\providecommand \@@startlink[1]{}%
\providecommand \@@endlink[0]{}%
\providecommand \url  [0]{\begingroup\@sanitize@url \@url }%
\providecommand \@url [1]{\endgroup\@href {#1}{\urlprefix }}%
\providecommand \urlprefix  [0]{URL }%
\providecommand \Eprint [0]{\href }%
\providecommand \doibase [0]{http://dx.doi.org/}%
\providecommand \selectlanguage [0]{\@gobble}%
\providecommand \bibinfo  [0]{\@secondoftwo}%
\providecommand \bibfield  [0]{\@secondoftwo}%
\providecommand \translation [1]{[#1]}%
\providecommand \BibitemOpen [0]{}%
\providecommand \bibitemStop [0]{}%
\providecommand \bibitemNoStop [0]{.\EOS\space}%
\providecommand \EOS [0]{\spacefactor3000\relax}%
\providecommand \BibitemShut  [1]{\csname bibitem#1\endcsname}%
\let\auto@bib@innerbib\@empty
%</preamble>
\bibitem [{\citenamefont {Tully}(2012)}]{Tully2012perspective}%
  \BibitemOpen
  \bibfield  {author} {\bibinfo {author} {\bibfnamefont {J.~C.}\ \bibnamefont
  {Tully}},\ }\href {\doibase 10.1063/1.4757762} {\bibfield  {journal}
  {\bibinfo  {journal} {J.~Chem. Phys.}\ }\textbf {\bibinfo {volume} {137}},\
  \bibinfo {pages} {22A301} (\bibinfo {year} {2012})}\BibitemShut {NoStop}%
\bibitem [{\citenamefont {Domcke}, \citenamefont {Yarkony},\ and\ \citenamefont
  {K\"oppel}(2004)}]{ConicalIntersections1}%
  \BibitemOpen
  \bibinfo {editor} {\bibfnamefont {W.}~\bibnamefont {Domcke}}, \bibinfo
  {editor} {\bibfnamefont {D.~R.}\ \bibnamefont {Yarkony}}, \ and\ \bibinfo
  {editor} {\bibfnamefont {H.}~\bibnamefont {K\"oppel}},\ eds.,\ \href@noop {}
  {\emph {\bibinfo {title} {Conical Intersections: {E}lectronic Structure,
  Dynamics and Spectroscopy}}}\ (\bibinfo  {publisher} {World Scientific},\
  \bibinfo {address} {Singapore},\ \bibinfo {year} {2004})\BibitemShut
  {NoStop}%
\bibitem [{\citenamefont {Levine}\ and\ \citenamefont
  {Mart{\'\i}nez}(2007)}]{Levine2007AIMS}%
  \BibitemOpen
  \bibfield  {author} {\bibinfo {author} {\bibfnamefont {B.~G.}\ \bibnamefont
  {Levine}}\ and\ \bibinfo {author} {\bibfnamefont {T.~J.}\ \bibnamefont
  {Mart{\'\i}nez}},\ }\href {\doibase
  10.1146/annurev.physchem.57.032905.104612} {\bibfield  {journal} {\bibinfo
  {journal} {Annu. Rev. Phys. Chem.}\ }\textbf {\bibinfo {volume} {58}},\
  \bibinfo {pages} {613} (\bibinfo {year} {2007})}\BibitemShut {NoStop}%
\bibitem [{\citenamefont {Stock}\ and\ \citenamefont
  {Thoss}(2005)}]{Stock2005nonadiabatic}%
  \BibitemOpen
  \bibfield  {author} {\bibinfo {author} {\bibfnamefont {G.}~\bibnamefont
  {Stock}}\ and\ \bibinfo {author} {\bibfnamefont {M.}~\bibnamefont {Thoss}},\
  }\href@noop {} {\bibfield  {journal} {\bibinfo  {journal} {Adv. Chem. Phys.}\
  }\textbf {\bibinfo {volume} {131}},\ \bibinfo {pages} {243} (\bibinfo {year}
  {2005})}\BibitemShut {NoStop}%
\bibitem [{\citenamefont {Garg}, \citenamefont {Onuchic},\ and\ \citenamefont
  {Ambegaokar}(1985)}]{Garg1985spinboson}%
  \BibitemOpen
  \bibfield  {author} {\bibinfo {author} {\bibfnamefont {A.}~\bibnamefont
  {Garg}}, \bibinfo {author} {\bibfnamefont {J.~N.}\ \bibnamefont {Onuchic}}, \
  and\ \bibinfo {author} {\bibfnamefont {V.}~\bibnamefont {Ambegaokar}},\
  }\href {\doibase 10.1063/1.449017} {\bibfield  {journal} {\bibinfo  {journal}
  {J.~Chem. Phys.}\ }\textbf {\bibinfo {volume} {83}},\ \bibinfo {pages} {4491}
  (\bibinfo {year} {1985})}\BibitemShut {NoStop}%
\bibitem [{\citenamefont {Leggett}\ \emph {et~al.}(1987)\citenamefont
  {Leggett}, \citenamefont {Chakravarty}, \citenamefont {Dorsey}, \citenamefont
  {Fisher}, \citenamefont {Garg},\ and\ \citenamefont
  {Zwerger}}]{Leggett1987spinboson}%
  \BibitemOpen
  \bibfield  {author} {\bibinfo {author} {\bibfnamefont {A.~J.}\ \bibnamefont
  {Leggett}}, \bibinfo {author} {\bibfnamefont {S.}~\bibnamefont
  {Chakravarty}}, \bibinfo {author} {\bibfnamefont {A.~T.}\ \bibnamefont
  {Dorsey}}, \bibinfo {author} {\bibfnamefont {M.~P.~A.}\ \bibnamefont
  {Fisher}}, \bibinfo {author} {\bibfnamefont {A.}~\bibnamefont {Garg}}, \ and\
  \bibinfo {author} {\bibfnamefont {W.}~\bibnamefont {Zwerger}},\ }\href
  {\doibase 10.1103/RevModPhys.59.1} {\bibfield  {journal} {\bibinfo  {journal}
  {Rev. Mod. Phys.}\ }\textbf {\bibinfo {volume} {59}},\ \bibinfo {pages} {1}
  (\bibinfo {year} {1987})}\BibitemShut {NoStop}%
\bibitem [{\citenamefont {Rips}\ and\ \citenamefont
  {Pollak}(1995)}]{Rips1995ET}%
  \BibitemOpen
  \bibfield  {author} {\bibinfo {author} {\bibfnamefont {I.}~\bibnamefont
  {Rips}}\ and\ \bibinfo {author} {\bibfnamefont {E.}~\bibnamefont {Pollak}},\
  }\href {\doibase 10.1063/1.470209} {\bibfield  {journal} {\bibinfo  {journal}
  {J.~Chem. Phys.}\ }\textbf {\bibinfo {volume} {103}},\ \bibinfo {pages}
  {7912} (\bibinfo {year} {1995})}\BibitemShut {NoStop}%
\bibitem [{\citenamefont {Weiss}(2008)}]{Weiss}%
  \BibitemOpen
  \bibfield  {author} {\bibinfo {author} {\bibfnamefont {U.}~\bibnamefont
  {Weiss}},\ }\href@noop {} {\emph {\bibinfo {title} {Quantum Dissipative
  Systems}}},\ \bibinfo {edition} {3rd}\ ed.\ (\bibinfo  {publisher} {World
  Scientific},\ \bibinfo {address} {Sinagpore},\ \bibinfo {year}
  {2008})\BibitemShut {NoStop}%
\bibitem [{\citenamefont {Wang}, \citenamefont {Thoss},\ and\ \citenamefont
  {Miller}(2001)}]{Wang2001hybrid}%
  \BibitemOpen
  \bibfield  {author} {\bibinfo {author} {\bibfnamefont {H.}~\bibnamefont
  {Wang}}, \bibinfo {author} {\bibfnamefont {M.}~\bibnamefont {Thoss}}, \ and\
  \bibinfo {author} {\bibfnamefont {W.~H.}\ \bibnamefont {Miller}},\ }\href
  {\doibase 10.1063/1.1385561} {\bibfield  {journal} {\bibinfo  {journal}
  {J.~Chem. Phys.}\ }\textbf {\bibinfo {volume} {115}},\ \bibinfo {pages}
  {2979} (\bibinfo {year} {2001})}\BibitemShut {NoStop}%
\bibitem [{\citenamefont {Thoss}, \citenamefont {Wang},\ and\ \citenamefont
  {Miller}(2001)}]{Thoss2001hybrid}%
  \BibitemOpen
  \bibfield  {author} {\bibinfo {author} {\bibfnamefont {M.}~\bibnamefont
  {Thoss}}, \bibinfo {author} {\bibfnamefont {H.}~\bibnamefont {Wang}}, \ and\
  \bibinfo {author} {\bibfnamefont {W.~H.}\ \bibnamefont {Miller}},\ }\href
  {\doibase 10.1063/1.1385562} {\bibfield  {journal} {\bibinfo  {journal}
  {J.~Chem. Phys.}\ }\textbf {\bibinfo {volume} {115}},\ \bibinfo {pages}
  {2991} (\bibinfo {year} {2001})}\BibitemShut {NoStop}%
\bibitem [{\citenamefont {Wang}\ and\ \citenamefont
  {Thoss}(2003)}]{Wang2003MLMCTDH}%
  \BibitemOpen
  \bibfield  {author} {\bibinfo {author} {\bibfnamefont {H.}~\bibnamefont
  {Wang}}\ and\ \bibinfo {author} {\bibfnamefont {M.}~\bibnamefont {Thoss}},\
  }\href {\doibase 10.1063/1.1580111} {\bibfield  {journal} {\bibinfo
  {journal} {J.~Chem. Phys.}\ }\textbf {\bibinfo {volume} {119}},\ \bibinfo
  {pages} {1289} (\bibinfo {year} {2003})}\BibitemShut {NoStop}%
\bibitem [{\citenamefont {Thoss}\ and\ \citenamefont
  {Wang}(2006)}]{Thoss2006MLMCTDH}%
  \BibitemOpen
  \bibfield  {author} {\bibinfo {author} {\bibfnamefont {M.}~\bibnamefont
  {Thoss}}\ and\ \bibinfo {author} {\bibfnamefont {H.}~\bibnamefont {Wang}},\
  }\href {\doibase 10.1016/j.chemphys.2005.07.011} {\bibfield  {journal}
  {\bibinfo  {journal} {Chem. Phys.}\ }\textbf {\bibinfo {volume} {322}},\
  \bibinfo {pages} {210} (\bibinfo {year} {2006})}\BibitemShut {NoStop}%
\bibitem [{\citenamefont {Wang}, \citenamefont {Skinner},\ and\ \citenamefont
  {Thoss}(2006)}]{Wang2006flux}%
  \BibitemOpen
  \bibfield  {author} {\bibinfo {author} {\bibfnamefont {H.}~\bibnamefont
  {Wang}}, \bibinfo {author} {\bibfnamefont {D.~E.}\ \bibnamefont {Skinner}}, \
  and\ \bibinfo {author} {\bibfnamefont {M.}~\bibnamefont {Thoss}},\ }\href
  {\doibase 10.1063/1.2363195} {\bibfield  {journal} {\bibinfo  {journal}
  {J.~Chem. Phys.}\ }\textbf {\bibinfo {volume} {125}},\ \bibinfo {pages}
  {174502} (\bibinfo {year} {2006})}\BibitemShut {NoStop}%
\bibitem [{\citenamefont {Mak}\ and\ \citenamefont
  {Chandler}(1991)}]{Mak1991spinboson}%
  \BibitemOpen
  \bibfield  {author} {\bibinfo {author} {\bibfnamefont {C.~H.}\ \bibnamefont
  {Mak}}\ and\ \bibinfo {author} {\bibfnamefont {D.}~\bibnamefont {Chandler}},\
  }\href {\doibase 10.1103/PhysRevA.44.2352} {\bibfield  {journal} {\bibinfo
  {journal} {Phys. Rev. A}\ }\textbf {\bibinfo {volume} {44}},\ \bibinfo
  {pages} {2352} (\bibinfo {year} {1991})}\BibitemShut {NoStop}%
\bibitem [{\citenamefont {Topaler}\ and\ \citenamefont
  {Makri}(1996)}]{Topaler1996nonadiabatic}%
  \BibitemOpen
  \bibfield  {author} {\bibinfo {author} {\bibfnamefont {M.}~\bibnamefont
  {Topaler}}\ and\ \bibinfo {author} {\bibfnamefont {N.}~\bibnamefont
  {Makri}},\ }\href {\doibase 10.1021/jp951673k} {\bibfield  {journal}
  {\bibinfo  {journal} {J.~Phys. Chem.}\ }\textbf {\bibinfo {volume} {100}},\
  \bibinfo {pages} {4430} (\bibinfo {year} {1996})}\BibitemShut {NoStop}%
\bibitem [{\citenamefont {M{\"u}hlbacher}\ and\ \citenamefont
  {Egger}(2003)}]{Muehlbacher2003spinboson}%
  \BibitemOpen
  \bibfield  {author} {\bibinfo {author} {\bibfnamefont {L.}~\bibnamefont
  {M{\"u}hlbacher}}\ and\ \bibinfo {author} {\bibfnamefont {R.}~\bibnamefont
  {Egger}},\ }\href {\doibase 10.1063/1.1523014} {\bibfield  {journal}
  {\bibinfo  {journal} {J.~Chem. Phys.}\ }\textbf {\bibinfo {volume} {118}},\
  \bibinfo {pages} {179} (\bibinfo {year} {2003})}\BibitemShut {NoStop}%
\bibitem [{\citenamefont {M{\"u}hlbacher}\ and\ \citenamefont
  {Egger}(2004)}]{Muehlbacher2004asymmetric}%
  \BibitemOpen
  \bibfield  {author} {\bibinfo {author} {\bibfnamefont {L.}~\bibnamefont
  {M{\"u}hlbacher}}\ and\ \bibinfo {author} {\bibfnamefont {R.}~\bibnamefont
  {Egger}},\ }\href {\doibase 10.1016/j.chemphys.2003.08.021} {\bibfield
  {journal} {\bibinfo  {journal} {Chem. Phys.}\ }\textbf {\bibinfo {volume}
  {296}},\ \bibinfo {pages} {193} (\bibinfo {year} {2004})}\BibitemShut
  {NoStop}%
\bibitem [{\citenamefont {Zwanzig}(2001)}]{Zwanzig}%
  \BibitemOpen
  \bibfield  {author} {\bibinfo {author} {\bibfnamefont {R.}~\bibnamefont
  {Zwanzig}},\ }\href@noop {} {\emph {\bibinfo {title} {Nonequilibrium
  Statistical Mechanics}}}\ (\bibinfo  {publisher} {Oxford University Press},\
  \bibinfo {year} {2001})\BibitemShut {NoStop}%
\bibitem [{\citenamefont {Lee}, \citenamefont {Dunietz},\ and\ \citenamefont
  {Geva}(2013)}]{Lee2013goldenrule}%
  \BibitemOpen
  \bibfield  {author} {\bibinfo {author} {\bibfnamefont {M.~H.}\ \bibnamefont
  {Lee}}, \bibinfo {author} {\bibfnamefont {B.~D.}\ \bibnamefont {Dunietz}}, \
  and\ \bibinfo {author} {\bibfnamefont {E.}~\bibnamefont {Geva}},\ }\href
  {\doibase 10.1021/jp4081417} {\bibfield  {journal} {\bibinfo  {journal}
  {J.~Phys. Chem.~C}\ }\textbf {\bibinfo {volume} {117}},\ \bibinfo {pages}
  {23391} (\bibinfo {year} {2013})}\BibitemShut {NoStop}%
\bibitem [{\citenamefont {Bader}, \citenamefont {Kuharski},\ and\ \citenamefont
  {Chandler}(1990)}]{Bader1990golden}%
  \BibitemOpen
  \bibfield  {author} {\bibinfo {author} {\bibfnamefont {J.~S.}\ \bibnamefont
  {Bader}}, \bibinfo {author} {\bibfnamefont {R.~A.}\ \bibnamefont {Kuharski}},
  \ and\ \bibinfo {author} {\bibfnamefont {D.}~\bibnamefont {Chandler}},\
  }\href {\doibase http://dx.doi.org/10.1063/1.459596} {\bibfield  {journal}
  {\bibinfo  {journal} {J.~Chem. Phys.}\ }\textbf {\bibinfo {volume} {93}},\
  \bibinfo {pages} {230} (\bibinfo {year} {1990})}\BibitemShut {NoStop}%
\bibitem [{\citenamefont {Marcus}(1956)}]{Marcus1956ET}%
  \BibitemOpen
  \bibfield  {author} {\bibinfo {author} {\bibfnamefont {R.~A.}\ \bibnamefont
  {Marcus}},\ }\href {\doibase 10.1063/1.1742723} {\bibfield  {journal}
  {\bibinfo  {journal} {J.~Chem. Phys.}\ }\textbf {\bibinfo {volume} {24}},\
  \bibinfo {pages} {966} (\bibinfo {year} {1956})}\BibitemShut {NoStop}%
\bibitem [{\citenamefont {Chandler}(1998)}]{ChandlerET}%
  \BibitemOpen
  \bibfield  {author} {\bibinfo {author} {\bibfnamefont {D.}~\bibnamefont
  {Chandler}},\ }in\ \href@noop {} {\emph {\bibinfo {booktitle} {Classical and
  Quantum Dynamics in Condensed Phase Simulations}}},\ \bibinfo {editor}
  {edited by\ \bibinfo {editor} {\bibfnamefont {B.~J.}\ \bibnamefont {Berne}},
  \bibinfo {editor} {\bibfnamefont {G.}~\bibnamefont {Ciccotti}}, \ and\
  \bibinfo {editor} {\bibfnamefont {D.~F.}\ \bibnamefont {Coker}}}\ (\bibinfo
  {publisher} {World Scientific},\ \bibinfo {address} {Singapore},\ \bibinfo
  {year} {1998})\ Chap.~\bibinfo {chapter} {2}, pp.\ \bibinfo {pages}
  {25--49}\BibitemShut {NoStop}%
\bibitem [{\citenamefont {Miller}, \citenamefont {Calcaterra},\ and\
  \citenamefont {Closs}(1984)}]{Miller1984inverted}%
  \BibitemOpen
  \bibfield  {author} {\bibinfo {author} {\bibfnamefont {J.~R.}\ \bibnamefont
  {Miller}}, \bibinfo {author} {\bibfnamefont {L.~T.}\ \bibnamefont
  {Calcaterra}}, \ and\ \bibinfo {author} {\bibfnamefont {G.~L.}\ \bibnamefont
  {Closs}},\ }\href {\doibase 10.1021/ja00322a058} {\bibfield  {journal}
  {\bibinfo  {journal} {J. Am. Chem. Soc.}\ }\textbf {\bibinfo {volume}
  {106}},\ \bibinfo {pages} {3047} (\bibinfo {year} {1984})}\BibitemShut
  {NoStop}%
\bibitem [{\citenamefont {Egorov}, \citenamefont {Rabani},\ and\ \citenamefont
  {Berne}(1999)}]{Egorov1999goldenrule}%
  \BibitemOpen
  \bibfield  {author} {\bibinfo {author} {\bibfnamefont {S.~A.}\ \bibnamefont
  {Egorov}}, \bibinfo {author} {\bibfnamefont {E.}~\bibnamefont {Rabani}}, \
  and\ \bibinfo {author} {\bibfnamefont {B.}~\bibnamefont {Berne}},\ }\href
  {\doibase 10.1021/jp9921349} {\bibfield  {journal} {\bibinfo  {journal}
  {J.~Phys.\ Chem.~B}\ }\textbf {\bibinfo {volume} {103}},\ \bibinfo {pages}
  {10978} (\bibinfo {year} {1999})}\BibitemShut {NoStop}%
\bibitem [{\citenamefont {Shi}\ and\ \citenamefont
  {Geva}(2004)}]{Shi2004goldenrule}%
  \BibitemOpen
  \bibfield  {author} {\bibinfo {author} {\bibfnamefont {Q.}~\bibnamefont
  {Shi}}\ and\ \bibinfo {author} {\bibfnamefont {E.}~\bibnamefont {Geva}},\
  }\href {\doibase 10.1021/jp049547g} {\bibfield  {journal} {\bibinfo
  {journal} {J.~Phys.\ Chem.~A}\ }\textbf {\bibinfo {volume} {108}},\ \bibinfo
  {pages} {6109} (\bibinfo {year} {2004})}\BibitemShut {NoStop}%
\bibitem [{\citenamefont {Wolynes}(1987)}]{Wolynes1987nonadiabatic}%
  \BibitemOpen
  \bibfield  {author} {\bibinfo {author} {\bibfnamefont {P.~G.}\ \bibnamefont
  {Wolynes}},\ }\href {\doibase 10.1063/1.453440} {\bibfield  {journal}
  {\bibinfo  {journal} {J.~Chem. Phys.}\ }\textbf {\bibinfo {volume} {87}},\
  \bibinfo {pages} {6559} (\bibinfo {year} {1987})}\BibitemShut {NoStop}%
\bibitem [{\citenamefont {Cao}\ and\ \citenamefont
  {Voth}(1997)}]{Cao1997nonadiabatic}%
  \BibitemOpen
  \bibfield  {author} {\bibinfo {author} {\bibfnamefont {J.}~\bibnamefont
  {Cao}}\ and\ \bibinfo {author} {\bibfnamefont {G.~A.}\ \bibnamefont {Voth}},\
  }\href {\doibase 10.1063/1.474123} {\bibfield  {journal} {\bibinfo  {journal}
  {J.~Chem. Phys.}\ }\textbf {\bibinfo {volume} {106}},\ \bibinfo {pages}
  {1769} (\bibinfo {year} {1997})}\BibitemShut {NoStop}%
\bibitem [{\citenamefont {Coleman}(1977)}]{Coleman1977ImF}%
  \BibitemOpen
  \bibfield  {author} {\bibinfo {author} {\bibfnamefont {S.}~\bibnamefont
  {Coleman}},\ }\href {\doibase 10.1103/PhysRevD.15.2929} {\bibfield  {journal}
  {\bibinfo  {journal} {Phys. Rev.~D}\ }\textbf {\bibinfo {volume} {15}},\
  \bibinfo {pages} {2929} (\bibinfo {year} {1977})}\BibitemShut {NoStop}%
\bibitem [{\citenamefont {Affleck}(1981)}]{Affleck1981ImF}%
  \BibitemOpen
  \bibfield  {author} {\bibinfo {author} {\bibfnamefont {I.}~\bibnamefont
  {Affleck}},\ }\href {\doibase 10.1103/PhysRevLett.46.388} {\bibfield
  {journal} {\bibinfo  {journal} {Phys. Rev. Lett.}\ }\textbf {\bibinfo
  {volume} {46}},\ \bibinfo {pages} {388} (\bibinfo {year} {1981})}\BibitemShut
  {NoStop}%
\bibitem [{\citenamefont {Zheng}, \citenamefont {McCammon},\ and\ \citenamefont
  {Wolynes}(1989)}]{Zheng1989ET}%
  \BibitemOpen
  \bibfield  {author} {\bibinfo {author} {\bibfnamefont {C.}~\bibnamefont
  {Zheng}}, \bibinfo {author} {\bibfnamefont {J.~A.}\ \bibnamefont {McCammon}},
  \ and\ \bibinfo {author} {\bibfnamefont {P.~G.}\ \bibnamefont {Wolynes}},\
  }\href@noop {} {\bibfield  {journal} {\bibinfo  {journal} {P. Natl. Acad.
  Sci. USA}\ }\textbf {\bibinfo {volume} {86}},\ \bibinfo {pages} {6441}
  (\bibinfo {year} {1989})}\BibitemShut {NoStop}%
\bibitem [{\citenamefont {Zheng}, \citenamefont {McCammon},\ and\ \citenamefont
  {Wolynes}(1991)}]{Zheng1991ET}%
  \BibitemOpen
  \bibfield  {author} {\bibinfo {author} {\bibfnamefont {C.}~\bibnamefont
  {Zheng}}, \bibinfo {author} {\bibfnamefont {J.~A.}\ \bibnamefont {McCammon}},
  \ and\ \bibinfo {author} {\bibfnamefont {P.~G.}\ \bibnamefont {Wolynes}},\
  }\href {\doibase 10.1016/0301-0104(91)87070-C} {\bibfield  {journal}
  {\bibinfo  {journal} {Chem. Phys.}\ }\textbf {\bibinfo {volume} {158}},\
  \bibinfo {pages} {261} (\bibinfo {year} {1991})}\BibitemShut {NoStop}%
\bibitem [{\citenamefont {Marchi}\ and\ \citenamefont
  {Chandler}(1991)}]{Marchi1991tunnelling}%
  \BibitemOpen
  \bibfield  {author} {\bibinfo {author} {\bibfnamefont {M.}~\bibnamefont
  {Marchi}}\ and\ \bibinfo {author} {\bibfnamefont {D.}~\bibnamefont
  {Chandler}},\ }\href {\doibase 10.1063/1.461096} {\bibfield  {journal}
  {\bibinfo  {journal} {J.~Chem. Phys.}\ }\textbf {\bibinfo {volume} {95}},\
  \bibinfo {pages} {889} (\bibinfo {year} {1991})}\BibitemShut {NoStop}%
\bibitem [{\citenamefont {Schwieters}\ and\ \citenamefont
  {Voth}(1999)}]{Schwieters1999diabatic}%
  \BibitemOpen
  \bibfield  {author} {\bibinfo {author} {\bibfnamefont {C.~D.}\ \bibnamefont
  {Schwieters}}\ and\ \bibinfo {author} {\bibfnamefont {G.~A.}\ \bibnamefont
  {Voth}},\ }\href {\doibase 10.1063/1.479569} {\bibfield  {journal} {\bibinfo
  {journal} {J.~Chem. Phys.}\ }\textbf {\bibinfo {volume} {111}},\ \bibinfo
  {pages} {2869} (\bibinfo {year} {1999})}\BibitemShut {NoStop}%
\bibitem [{\citenamefont {Miller}, \citenamefont {Schwartz},\ and\
  \citenamefont {Tromp}(1983)}]{Miller1983rate}%
  \BibitemOpen
  \bibfield  {author} {\bibinfo {author} {\bibfnamefont {W.~H.}\ \bibnamefont
  {Miller}}, \bibinfo {author} {\bibfnamefont {S.~D.}\ \bibnamefont
  {Schwartz}}, \ and\ \bibinfo {author} {\bibfnamefont {J.~W.}\ \bibnamefont
  {Tromp}},\ }\href {\doibase 10.1063/1.445581} {\bibfield  {journal} {\bibinfo
   {journal} {J.~Chem. Phys.}\ }\textbf {\bibinfo {volume} {79}},\ \bibinfo
  {pages} {4889} (\bibinfo {year} {1983})}\BibitemShut {NoStop}%
\bibitem [{\citenamefont {Tully}(1990)}]{Tully1990hopping}%
  \BibitemOpen
  \bibfield  {author} {\bibinfo {author} {\bibfnamefont {J.~C.}\ \bibnamefont
  {Tully}},\ }\href {\doibase 10.1063/1.459170} {\bibfield  {journal} {\bibinfo
   {journal} {J.~Chem. Phys.}\ }\textbf {\bibinfo {volume} {93}},\ \bibinfo
  {pages} {1061} (\bibinfo {year} {1990})}\BibitemShut {NoStop}%
\bibitem [{\citenamefont {Landry}\ and\ \citenamefont
  {Subotnik}(2011)}]{Landry2011hopping}%
  \BibitemOpen
  \bibfield  {author} {\bibinfo {author} {\bibfnamefont {B.~R.}\ \bibnamefont
  {Landry}}\ and\ \bibinfo {author} {\bibfnamefont {J.~E.}\ \bibnamefont
  {Subotnik}},\ }\href {\doibase 10.1063/1.3663870} {\bibfield  {journal}
  {\bibinfo  {journal} {J.~Chem. Phys.}\ }\textbf {\bibinfo {volume} {135}},\
  \bibinfo {pages} {191101} (\bibinfo {year} {2011})}\BibitemShut {NoStop}%
\bibitem [{\citenamefont {Landry}\ and\ \citenamefont
  {Subotnik}(2012)}]{Landry2012hopping}%
  \BibitemOpen
  \bibfield  {author} {\bibinfo {author} {\bibfnamefont {B.~R.}\ \bibnamefont
  {Landry}}\ and\ \bibinfo {author} {\bibfnamefont {J.~E.}\ \bibnamefont
  {Subotnik}},\ }\href {\doibase 10.1063/1.4733675} {\bibfield  {journal}
  {\bibinfo  {journal} {J.~Chem. Phys.}\ }\textbf {\bibinfo {volume} {137}},\
  \bibinfo {pages} {22A513} (\bibinfo {year} {2012})}\BibitemShut {NoStop}%
\bibitem [{\citenamefont {Meyer}\ and\ \citenamefont
  {Miller}(1979)}]{Meyer1979nonadiabatic}%
  \BibitemOpen
  \bibfield  {author} {\bibinfo {author} {\bibfnamefont {H.-D.}\ \bibnamefont
  {Meyer}}\ and\ \bibinfo {author} {\bibfnamefont {W.~H.}\ \bibnamefont
  {Miller}},\ }\href {\doibase 10.1063/1.437910} {\bibfield  {journal}
  {\bibinfo  {journal} {J.~Chem. Phys.}\ }\textbf {\bibinfo {volume} {70}},\
  \bibinfo {pages} {3214} (\bibinfo {year} {1979})}\BibitemShut {NoStop}%
\bibitem [{\citenamefont {Stock}\ and\ \citenamefont
  {Thoss}(1997)}]{Stock1997mapping}%
  \BibitemOpen
  \bibfield  {author} {\bibinfo {author} {\bibfnamefont {G.}~\bibnamefont
  {Stock}}\ and\ \bibinfo {author} {\bibfnamefont {M.}~\bibnamefont {Thoss}},\
  }\href {\doibase 10.1103/PhysRevLett.78.578} {\bibfield  {journal} {\bibinfo
  {journal} {Phys. Rev. Lett.}\ }\textbf {\bibinfo {volume} {78}},\ \bibinfo
  {pages} {578} (\bibinfo {year} {1997})}\BibitemShut {NoStop}%
\bibitem [{\citenamefont {Thoss}\ and\ \citenamefont
  {Stock}(1999)}]{Thoss1999mapping}%
  \BibitemOpen
  \bibfield  {author} {\bibinfo {author} {\bibfnamefont {M.}~\bibnamefont
  {Thoss}}\ and\ \bibinfo {author} {\bibfnamefont {G.}~\bibnamefont {Stock}},\
  }\href {\doibase 10.1103/PhysRevA.59.64} {\bibfield  {journal} {\bibinfo
  {journal} {Phys. Rev. A}\ }\textbf {\bibinfo {volume} {59}},\ \bibinfo
  {pages} {64} (\bibinfo {year} {1999})}\BibitemShut {NoStop}%
\bibitem [{\citenamefont {M{\"u}ller}\ and\ \citenamefont
  {Stock}(1998)}]{Mueller1998mapping}%
  \BibitemOpen
  \bibfield  {author} {\bibinfo {author} {\bibfnamefont {U.}~\bibnamefont
  {M{\"u}ller}}\ and\ \bibinfo {author} {\bibfnamefont {G.}~\bibnamefont
  {Stock}},\ }\href {\doibase 10.1063/1.476184} {\bibfield  {journal} {\bibinfo
   {journal} {J.~Chem. Phys.}\ }\textbf {\bibinfo {volume} {108}},\ \bibinfo
  {pages} {7516} (\bibinfo {year} {1998})}\BibitemShut {NoStop}%
\bibitem [{\citenamefont {M{\"u}ller}\ and\ \citenamefont
  {Stock}(1999)}]{Mueller1999pyrazine}%
  \BibitemOpen
  \bibfield  {author} {\bibinfo {author} {\bibfnamefont {U.}~\bibnamefont
  {M{\"u}ller}}\ and\ \bibinfo {author} {\bibfnamefont {G.}~\bibnamefont
  {Stock}},\ }\href {\doibase 10.1063/1.479255} {\bibfield  {journal} {\bibinfo
   {journal} {J.~Chem. Phys.}\ }\textbf {\bibinfo {volume} {111}},\ \bibinfo
  {pages} {77} (\bibinfo {year} {1999})}\BibitemShut {NoStop}%
\bibitem [{\citenamefont {Sun}\ and\ \citenamefont
  {Miller}(1997)}]{Sun1997mapping}%
  \BibitemOpen
  \bibfield  {author} {\bibinfo {author} {\bibfnamefont {X.}~\bibnamefont
  {Sun}}\ and\ \bibinfo {author} {\bibfnamefont {W.~H.}\ \bibnamefont
  {Miller}},\ }\href {\doibase 10.1063/1.473624} {\bibfield  {journal}
  {\bibinfo  {journal} {J.~Chem. Phys.}\ }\textbf {\bibinfo {volume} {106}},\
  \bibinfo {pages} {6346} (\bibinfo {year} {1997})}\BibitemShut {NoStop}%
\bibitem [{\citenamefont {Bonella}\ and\ \citenamefont
  {Coker}(2003)}]{Bonella2003mapping}%
  \BibitemOpen
  \bibfield  {author} {\bibinfo {author} {\bibfnamefont {S.}~\bibnamefont
  {Bonella}}\ and\ \bibinfo {author} {\bibfnamefont {D.~F.}\ \bibnamefont
  {Coker}},\ }\href {\doibase 10.1063/1.1542883} {\bibfield  {journal}
  {\bibinfo  {journal} {J.~Chem. Phys.}\ }\textbf {\bibinfo {volume} {118}},\
  \bibinfo {pages} {4370} (\bibinfo {year} {2003})}\BibitemShut {NoStop}%
\bibitem [{\citenamefont {Miller}(2009)}]{Miller2009mapping}%
  \BibitemOpen
  \bibfield  {author} {\bibinfo {author} {\bibfnamefont {W.~H.}\ \bibnamefont
  {Miller}},\ }\href {\doibase 10.1021/jp809907p} {\bibfield  {journal}
  {\bibinfo  {journal} {J.~Phys.\ Chem.~A}\ }\textbf {\bibinfo {volume}
  {113}},\ \bibinfo {pages} {1405} (\bibinfo {year} {2009})}\BibitemShut
  {NoStop}%
\bibitem [{\citenamefont {Sun}, \citenamefont {Wang},\ and\ \citenamefont
  {Miller}(1998)}]{Sun1998mapping}%
  \BibitemOpen
  \bibfield  {author} {\bibinfo {author} {\bibfnamefont {X.}~\bibnamefont
  {Sun}}, \bibinfo {author} {\bibfnamefont {H.}~\bibnamefont {Wang}}, \ and\
  \bibinfo {author} {\bibfnamefont {W.~H.}\ \bibnamefont {Miller}},\ }\href
  {\doibase 10.1063/1.477389} {\bibfield  {journal} {\bibinfo  {journal}
  {J.~Chem. Phys.}\ }\textbf {\bibinfo {volume} {109}},\ \bibinfo {pages}
  {7064} (\bibinfo {year} {1998})}\BibitemShut {NoStop}%
\bibitem [{\citenamefont {Wang}\ \emph {et~al.}(1999)\citenamefont {Wang},
  \citenamefont {Song}, \citenamefont {Chandler},\ and\ \citenamefont
  {Miller}}]{Wang1999mapping}%
  \BibitemOpen
  \bibfield  {author} {\bibinfo {author} {\bibfnamefont {H.}~\bibnamefont
  {Wang}}, \bibinfo {author} {\bibfnamefont {X.}~\bibnamefont {Song}}, \bibinfo
  {author} {\bibfnamefont {D.}~\bibnamefont {Chandler}}, \ and\ \bibinfo
  {author} {\bibfnamefont {W.~H.}\ \bibnamefont {Miller}},\ }\href {\doibase
  10.1063/1.478388} {\bibfield  {journal} {\bibinfo  {journal} {J.~Chem.
  Phys.}\ }\textbf {\bibinfo {volume} {110}},\ \bibinfo {pages} {4828}
  (\bibinfo {year} {1999})}\BibitemShut {NoStop}%
\bibitem [{\citenamefont {Liao}\ and\ \citenamefont
  {Voth}(2002)}]{Liao2002mappingCMD}%
  \BibitemOpen
  \bibfield  {author} {\bibinfo {author} {\bibfnamefont {J.-L.}\ \bibnamefont
  {Liao}}\ and\ \bibinfo {author} {\bibfnamefont {G.~A.}\ \bibnamefont
  {Voth}},\ }\href {\doibase 10.1021/jp020978d} {\bibfield  {journal} {\bibinfo
   {journal} {J.~Phys.\ Chem.~B}\ }\textbf {\bibinfo {volume} {106}},\ \bibinfo
  {pages} {8449} (\bibinfo {year} {2002})}\BibitemShut {NoStop}%
\bibitem [{\citenamefont {Ananth}\ and\ \citenamefont
  {Miller}(2010)}]{Ananth2010mapping}%
  \BibitemOpen
  \bibfield  {author} {\bibinfo {author} {\bibfnamefont {N.}~\bibnamefont
  {Ananth}}\ and\ \bibinfo {author} {\bibfnamefont {T.~F.}\ \bibnamefont
  {Miller}, \bibfnamefont {III}},\ }\href {\doibase 10.1063/1.3511700}
  {\bibfield  {journal} {\bibinfo  {journal} {J.~Chem. Phys.}\ }\textbf
  {\bibinfo {volume} {133}},\ \bibinfo {pages} {234103} (\bibinfo {year}
  {2010})}\BibitemShut {NoStop}%
\bibitem [{\citenamefont {Richardson}\ and\ \citenamefont
  {Thoss}(2013)}]{mapping}%
  \BibitemOpen
  \bibfield  {author} {\bibinfo {author} {\bibfnamefont {J.~O.}\ \bibnamefont
  {Richardson}}\ and\ \bibinfo {author} {\bibfnamefont {M.}~\bibnamefont
  {Thoss}},\ }\href {\doibase 10.1063/1.4816124} {\bibfield  {journal}
  {\bibinfo  {journal} {J.~Chem. Phys.}\ }\textbf {\bibinfo {volume} {139}},\
  \bibinfo {pages} {031102} (\bibinfo {year} {2013})}\BibitemShut {NoStop}%
\bibitem [{\citenamefont {Ananth}(2013)}]{Ananth2013MVRPMD}%
  \BibitemOpen
  \bibfield  {author} {\bibinfo {author} {\bibfnamefont {N.}~\bibnamefont
  {Ananth}},\ }\href {\doibase 10.1063/1.4821590} {\bibfield  {journal}
  {\bibinfo  {journal} {J.~Chem. Phys.}\ }\textbf {\bibinfo {volume} {139}},\
  \bibinfo {pages} {124102} (\bibinfo {year} {2013})}\BibitemShut {NoStop}%
\bibitem [{\citenamefont {Huo}, \citenamefont {Miller~III},\ and\ \citenamefont
  {Coker}(2013)}]{Huo2013PLDM}%
  \BibitemOpen
  \bibfield  {author} {\bibinfo {author} {\bibfnamefont {P.}~\bibnamefont
  {Huo}}, \bibinfo {author} {\bibfnamefont {T.~F.}\ \bibnamefont {Miller~III}},
  \ and\ \bibinfo {author} {\bibfnamefont {D.~F.}\ \bibnamefont {Coker}},\
  }\href {\doibase 10.1063/1.4826163} {\bibfield  {journal} {\bibinfo
  {journal} {J.~Chem. Phys.}\ }\textbf {\bibinfo {volume} {139}},\ \bibinfo
  {pages} {151103} (\bibinfo {year} {2013})}\BibitemShut {NoStop}%
\bibitem [{\citenamefont {Micha}(1983)}]{Micha1983Ehrenfest}%
  \BibitemOpen
  \bibfield  {author} {\bibinfo {author} {\bibfnamefont {D.~A.}\ \bibnamefont
  {Micha}},\ }\href {\doibase 10.1063/1.444753} {\bibfield  {journal} {\bibinfo
   {journal} {J.~Chem. Phys.}\ }\textbf {\bibinfo {volume} {78}},\ \bibinfo
  {pages} {7138} (\bibinfo {year} {1983})}\BibitemShut {NoStop}%
\bibitem [{\citenamefont {Ben-Nun}\ and\ \citenamefont
  {Mart{\'i}nez}(2002)}]{BenNun2002AIMS}%
  \BibitemOpen
  \bibfield  {author} {\bibinfo {author} {\bibfnamefont {M.}~\bibnamefont
  {Ben-Nun}}\ and\ \bibinfo {author} {\bibfnamefont {T.~J.}\ \bibnamefont
  {Mart{\'i}nez}},\ }\href@noop {} {\bibfield  {journal} {\bibinfo  {journal}
  {Adv. Chem. Phys.}\ }\textbf {\bibinfo {volume} {121}},\ \bibinfo {pages}
  {439} (\bibinfo {year} {2002})}\BibitemShut {NoStop}%
\bibitem [{\citenamefont {Kapral}(2006)}]{Kapral2006QCL}%
  \BibitemOpen
  \bibfield  {author} {\bibinfo {author} {\bibfnamefont {R.}~\bibnamefont
  {Kapral}},\ }\href {\doibase 10.1146/annurev.physchem.57.032905.104702}
  {\bibfield  {journal} {\bibinfo  {journal} {Annu. Rev. Phys. Chem.}\ }\textbf
  {\bibinfo {volume} {57}},\ \bibinfo {pages} {129} (\bibinfo {year}
  {2006})}\BibitemShut {NoStop}%
\bibitem [{\citenamefont {Agostini}\ \emph {et~al.}(2013)\citenamefont
  {Agostini}, \citenamefont {Abedi}, \citenamefont {Suzuki},\ and\
  \citenamefont {Gross}}]{Agostini2013MQC}%
  \BibitemOpen
  \bibfield  {author} {\bibinfo {author} {\bibfnamefont {F.}~\bibnamefont
  {Agostini}}, \bibinfo {author} {\bibfnamefont {A.}~\bibnamefont {Abedi}},
  \bibinfo {author} {\bibfnamefont {Y.}~\bibnamefont {Suzuki}}, \ and\ \bibinfo
  {author} {\bibfnamefont {E.}~\bibnamefont {Gross}},\ }\href {\doibase
  10.1080/00268976.2013.843731} {\bibfield  {journal} {\bibinfo  {journal}
  {Mol. Phys.}\ }\textbf {\bibinfo {volume} {111}},\ \bibinfo {pages} {3625}
  (\bibinfo {year} {2013})}\BibitemShut {NoStop}%
\bibitem [{\citenamefont {Menzeleev}, \citenamefont {Ananth},\ and\
  \citenamefont {Miller}(2011)}]{Menzeleev2011ET}%
  \BibitemOpen
  \bibfield  {author} {\bibinfo {author} {\bibfnamefont {A.~R.}\ \bibnamefont
  {Menzeleev}}, \bibinfo {author} {\bibfnamefont {N.}~\bibnamefont {Ananth}}, \
  and\ \bibinfo {author} {\bibfnamefont {T.~F.}\ \bibnamefont {Miller},
  \bibfnamefont {III}},\ }\href {\doibase 10.1063/1.3624766} {\bibfield
  {journal} {\bibinfo  {journal} {J.~Chem. Phys.}\ }\textbf {\bibinfo {volume}
  {135}},\ \bibinfo {pages} {074106} (\bibinfo {year} {2011})}\BibitemShut
  {NoStop}%
\bibitem [{\citenamefont {Kretchmer}\ and\ \citenamefont
  {Miller~III}(2013)}]{Kretchmer2013ET}%
  \BibitemOpen
  \bibfield  {author} {\bibinfo {author} {\bibfnamefont {J.~S.}\ \bibnamefont
  {Kretchmer}}\ and\ \bibinfo {author} {\bibfnamefont {T.~F.}\ \bibnamefont
  {Miller~III}},\ }\href {\doibase 10.1063/1.4797462} {\bibfield  {journal}
  {\bibinfo  {journal} {J.~Chem. Phys.}\ }\textbf {\bibinfo {volume} {138}},\
  \bibinfo {pages} {134109} (\bibinfo {year} {2013})}\BibitemShut {NoStop}%
\bibitem [{\citenamefont {Shushkov}(2013)}]{Shushkov2013instanton}%
  \BibitemOpen
  \bibfield  {author} {\bibinfo {author} {\bibfnamefont {P.}~\bibnamefont
  {Shushkov}},\ }\href@noop {} {\bibfield  {journal} {\bibinfo  {journal}
  {J.~Chem. Phys.}\ }\textbf {\bibinfo {volume} {138}},\ \bibinfo {pages}
  {224102} (\bibinfo {year} {2013})}\BibitemShut {NoStop}%
\bibitem [{\citenamefont {Menzeleev}, \citenamefont {Bell},\ and\ \citenamefont
  {Miller~III}(2014)}]{Menzeleev2014kinetic}%
  \BibitemOpen
  \bibfield  {author} {\bibinfo {author} {\bibfnamefont {A.~R.}\ \bibnamefont
  {Menzeleev}}, \bibinfo {author} {\bibfnamefont {F.}~\bibnamefont {Bell}}, \
  and\ \bibinfo {author} {\bibfnamefont {T.~F.}\ \bibnamefont {Miller~III}},\
  }\href {\doibase 10.1063/1.4863919} {\bibfield  {journal} {\bibinfo
  {journal} {J.~Chem. Phys.}\ }\textbf {\bibinfo {volume} {140}},\ \bibinfo
  {pages} {064103} (\bibinfo {year} {2014})}\BibitemShut {NoStop}%
\bibitem [{\citenamefont {Craig}\ and\ \citenamefont
  {Manolopoulos}(2005)}]{RPMDrefinedRate}%
  \BibitemOpen
  \bibfield  {author} {\bibinfo {author} {\bibfnamefont {I.~R.}\ \bibnamefont
  {Craig}}\ and\ \bibinfo {author} {\bibfnamefont {D.~E.}\ \bibnamefont
  {Manolopoulos}},\ }\href {\doibase 10.1063/1.1954769} {\bibfield  {journal}
  {\bibinfo  {journal} {J.~Chem. Phys.}\ }\textbf {\bibinfo {volume} {123}},\
  \bibinfo {pages} {034102} (\bibinfo {year} {2005})}\BibitemShut {NoStop}%
\bibitem [{\citenamefont {Richardson}\ and\ \citenamefont
  {Althorpe}(2009)}]{rpinst}%
  \BibitemOpen
  \bibfield  {author} {\bibinfo {author} {\bibfnamefont {J.~O.}\ \bibnamefont
  {Richardson}}\ and\ \bibinfo {author} {\bibfnamefont {S.~C.}\ \bibnamefont
  {Althorpe}},\ }\href {\doibase 10.1063/1.3267318} {\bibfield  {journal}
  {\bibinfo  {journal} {J.~Chem. Phys.}\ }\textbf {\bibinfo {volume} {131}},\
  \bibinfo {pages} {214106} (\bibinfo {year} {2009})}\BibitemShut {NoStop}%
\bibitem [{\citenamefont {Hele}\ and\ \citenamefont
  {Althorpe}(2013{\natexlab{a}})}]{Hele2013QTST}%
  \BibitemOpen
  \bibfield  {author} {\bibinfo {author} {\bibfnamefont {T.~J.~H.}\
  \bibnamefont {Hele}}\ and\ \bibinfo {author} {\bibfnamefont {S.~C.}\
  \bibnamefont {Althorpe}},\ }\href {\doibase 10.1063/1.4792697} {\bibfield
  {journal} {\bibinfo  {journal} {J.~Chem. Phys.}\ }\textbf {\bibinfo {volume}
  {138}},\ \bibinfo {pages} {084108} (\bibinfo {year}
  {2013}{\natexlab{a}})}\BibitemShut {NoStop}%
\bibitem [{\citenamefont {Althorpe}\ and\ \citenamefont
  {Hele}(2013)}]{Althorpe2013QTST}%
  \BibitemOpen
  \bibfield  {author} {\bibinfo {author} {\bibfnamefont {S.~C.}\ \bibnamefont
  {Althorpe}}\ and\ \bibinfo {author} {\bibfnamefont {T.~J.~H.}\ \bibnamefont
  {Hele}},\ }\href {\doibase 10.1063/1.4819076} {\bibfield  {journal} {\bibinfo
   {journal} {J.~Chem. Phys.}\ }\textbf {\bibinfo {volume} {139}},\ \bibinfo
  {pages} {084115} (\bibinfo {year} {2013})}\BibitemShut {NoStop}%
\bibitem [{\citenamefont {Hele}\ and\ \citenamefont
  {Althorpe}(2013{\natexlab{b}})}]{Hele2013unique}%
  \BibitemOpen
  \bibfield  {author} {\bibinfo {author} {\bibfnamefont {T.~J.~H.}\
  \bibnamefont {Hele}}\ and\ \bibinfo {author} {\bibfnamefont {S.~C.}\
  \bibnamefont {Althorpe}},\ }\href {\doibase 10.1063/1.4819077} {\bibfield
  {journal} {\bibinfo  {journal} {J.~Chem. Phys.}\ }\textbf {\bibinfo {volume}
  {139}},\ \bibinfo {pages} {084116} (\bibinfo {year}
  {2013}{\natexlab{b}})}\BibitemShut {NoStop}%
\bibitem [{\citenamefont {Chandler}(1987)}]{ChandlerGreen}%
  \BibitemOpen
  \bibfield  {author} {\bibinfo {author} {\bibfnamefont {D.}~\bibnamefont
  {Chandler}},\ }\href@noop {} {\emph {\bibinfo {title} {Introduction to Modern
  Statistical Mechanics}}}\ (\bibinfo  {publisher} {Oxford University Press},\
  \bibinfo {address} {New York},\ \bibinfo {year} {1987})\BibitemShut {NoStop}%
\bibitem [{\citenamefont {Chandler}(1978)}]{Chandler1978TST}%
  \BibitemOpen
  \bibfield  {author} {\bibinfo {author} {\bibfnamefont {D.}~\bibnamefont
  {Chandler}},\ }\href {\doibase 10.1063/1.436049} {\bibfield  {journal}
  {\bibinfo  {journal} {J.~Chem. Phys.}\ }\textbf {\bibinfo {volume} {68}},\
  \bibinfo {pages} {2959} (\bibinfo {year} {1978})}\BibitemShut {NoStop}%
\bibitem [{\citenamefont {Kubo}, \citenamefont {Toda},\ and\ \citenamefont
  {Hashitsume}(1978)}]{KuboBook}%
  \BibitemOpen
  \bibfield  {author} {\bibinfo {author} {\bibfnamefont {R.}~\bibnamefont
  {Kubo}}, \bibinfo {author} {\bibfnamefont {M.}~\bibnamefont {Toda}}, \ and\
  \bibinfo {author} {\bibfnamefont {N.}~\bibnamefont {Hashitsume}},\
  }\href@noop {} {\emph {\bibinfo {title} {Statistical Physics II}}}\ (\bibinfo
   {publisher} {Springer},\ \bibinfo {year} {1978})\BibitemShut {NoStop}%
\bibitem [{\citenamefont {Craig}, \citenamefont {Thoss},\ and\ \citenamefont
  {Wang}(2007)}]{Craig2007condensed}%
  \BibitemOpen
  \bibfield  {author} {\bibinfo {author} {\bibfnamefont {I.~R.}\ \bibnamefont
  {Craig}}, \bibinfo {author} {\bibfnamefont {M.}~\bibnamefont {Thoss}}, \ and\
  \bibinfo {author} {\bibfnamefont {H.}~\bibnamefont {Wang}},\ }\href {\doibase
  10.1063/1.2772265} {\bibfield  {journal} {\bibinfo  {journal} {J.~Chem.
  Phys.}\ }\textbf {\bibinfo {volume} {127}},\ \bibinfo {pages} {144503}
  (\bibinfo {year} {2007})}\BibitemShut {NoStop}%
\bibitem [{\citenamefont {Yamamoto}(1960)}]{Yamamoto1960rate}%
  \BibitemOpen
  \bibfield  {author} {\bibinfo {author} {\bibfnamefont {T.}~\bibnamefont
  {Yamamoto}},\ }\href {\doibase 10.1063/1.1731099} {\bibfield  {journal}
  {\bibinfo  {journal} {J.~Chem. Phys.}\ }\textbf {\bibinfo {volume} {33}},\
  \bibinfo {pages} {281} (\bibinfo {year} {1960})}\BibitemShut {NoStop}%
\bibitem [{\citenamefont {Miller}(1974)}]{Miller1974QTST}%
  \BibitemOpen
  \bibfield  {author} {\bibinfo {author} {\bibfnamefont {W.~H.}\ \bibnamefont
  {Miller}},\ }\href {\doibase 10.1063/1.1682181} {\bibfield  {journal}
  {\bibinfo  {journal} {J.~Chem. Phys.}\ }\textbf {\bibinfo {volume} {61}},\
  \bibinfo {pages} {1823} (\bibinfo {year} {1974})}\BibitemShut {NoStop}%
\bibitem [{\citenamefont {Craig}\ and\ \citenamefont
  {Manolopoulos}(2004)}]{RPMDcorrelation}%
  \BibitemOpen
  \bibfield  {author} {\bibinfo {author} {\bibfnamefont {I.~R.}\ \bibnamefont
  {Craig}}\ and\ \bibinfo {author} {\bibfnamefont {D.~E.}\ \bibnamefont
  {Manolopoulos}},\ }\href {\doibase 10.1063/1.1777575} {\bibfield  {journal}
  {\bibinfo  {journal} {J.~Chem. Phys.}\ }\textbf {\bibinfo {volume} {121}},\
  \bibinfo {pages} {3368} (\bibinfo {year} {2004})}\BibitemShut {NoStop}%
\bibitem [{\citenamefont {Borkovec}\ and\ \citenamefont
  {Talkner}(1990)}]{Borkovec1990rate}%
  \BibitemOpen
  \bibfield  {author} {\bibinfo {author} {\bibfnamefont {M.}~\bibnamefont
  {Borkovec}}\ and\ \bibinfo {author} {\bibfnamefont {P.}~\bibnamefont
  {Talkner}},\ }\href {\doibase 10.1063/1.458535} {\bibfield  {journal}
  {\bibinfo  {journal} {J.~Chem. Phys.}\ }\textbf {\bibinfo {volume} {92}},\
  \bibinfo {pages} {5307} (\bibinfo {year} {1990})}\BibitemShut {NoStop}%
\bibitem [{\citenamefont {Ruiz-Montero}, \citenamefont {Frenkel},\ and\
  \citenamefont {Brey}(1997)}]{Montero1997rate}%
  \BibitemOpen
  \bibfield  {author} {\bibinfo {author} {\bibfnamefont {M.~J.}\ \bibnamefont
  {Ruiz-Montero}}, \bibinfo {author} {\bibfnamefont {D.}~\bibnamefont
  {Frenkel}}, \ and\ \bibinfo {author} {\bibfnamefont {J.~J.}\ \bibnamefont
  {Brey}},\ }\href {\doibase 10.1080/002689797171922} {\bibfield  {journal}
  {\bibinfo  {journal} {Mol. Phys.}\ }\textbf {\bibinfo {volume} {90}},\
  \bibinfo {pages} {925} (\bibinfo {year} {1997})}\BibitemShut {NoStop}%
\bibitem [{\citenamefont {Frenkel}\ and\ \citenamefont {Smit}(1996)}]{MolSim}%
  \BibitemOpen
  \bibfield  {author} {\bibinfo {author} {\bibfnamefont {D.}~\bibnamefont
  {Frenkel}}\ and\ \bibinfo {author} {\bibfnamefont {B.}~\bibnamefont {Smit}},\
  }\href@noop {} {\emph {\bibinfo {title} {Understanding Molecular
  Simulation}}},\ \bibinfo {edition} {2nd}\ ed.\ (\bibinfo  {publisher}
  {Elsevier},\ \bibinfo {address} {San Diego},\ \bibinfo {year}
  {1996})\BibitemShut {NoStop}%
\bibitem [{\citenamefont {Zener}(1932)}]{Zener1932LZ}%
  \BibitemOpen
  \bibfield  {author} {\bibinfo {author} {\bibfnamefont {C.}~\bibnamefont
  {Zener}},\ }\href {\doibase 10.1098/rspa.1932.0165} {\bibfield  {journal}
  {\bibinfo  {journal} {Proc. R. Soc. Lond. A}\ }\textbf {\bibinfo {volume}
  {137}},\ \bibinfo {pages} {696} (\bibinfo {year} {1932})}\BibitemShut
  {NoStop}%
\bibitem [{\citenamefont {Nitzan}(2006)}]{Nitzan}%
  \BibitemOpen
  \bibfield  {author} {\bibinfo {author} {\bibfnamefont {A.}~\bibnamefont
  {Nitzan}},\ }\href@noop {} {\emph {\bibinfo {title} {Chemical Dynamics in
  Condensed Phases: Relaxation, Transfer, and Reactions in Condensed Molecular
  Systems}}}\ (\bibinfo  {publisher} {Oxford University Press},\ \bibinfo
  {year} {2006})\BibitemShut {NoStop}%
\bibitem [{Note1()}]{Note1}%
  \BibitemOpen
  \bibinfo {note} {As an extreme example, consider the case of a system where
  the two diabatic potentials have a minimum at the same nuclear configuration.
  A dividing surface would not then be able to differentiate between products
  and reactants.}\BibitemShut {Stop}%
\bibitem [{Note2()}]{Note2}%
  \BibitemOpen
  \bibinfo {note} {In fact, because of this it is not obvious if it is possible
  to find a single unified formula to give the rate constant for any value of
  $\Delta $ from the adiabatic to the nonadiabatic limit. There is no smooth
  connection as the very meaning and definition of the rate constant is
  different in these two regimes.}\BibitemShut {Stop}%
\bibitem [{Note3()}]{Note3}%
  \BibitemOpen
  \bibinfo {note} {In this case, the operator $B$ is the projection onto a
  diabatic state and not a side operator at all. Nonetheless we retain the
  familiar terminology for the flux-side correlation function.}\BibitemShut
  {Stop}%
\bibitem [{\citenamefont {H{\"a}nggi}, \citenamefont {Talkner},\ and\
  \citenamefont {Borkovec}(1990)}]{Haenggi1990rate}%
  \BibitemOpen
  \bibfield  {author} {\bibinfo {author} {\bibfnamefont {P.}~\bibnamefont
  {H{\"a}nggi}}, \bibinfo {author} {\bibfnamefont {P.}~\bibnamefont {Talkner}},
  \ and\ \bibinfo {author} {\bibfnamefont {M.}~\bibnamefont {Borkovec}},\
  }\href {\doibase 10.1103/RevModPhys.62.251} {\bibfield  {journal} {\bibinfo
  {journal} {Rev. Mod. Phys.}\ }\textbf {\bibinfo {volume} {62}},\ \bibinfo
  {pages} {251} (\bibinfo {year} {1990})}\BibitemShut {NoStop}%
\bibitem [{\citenamefont {Zimmermann}\ and\ \citenamefont
  {Van{\'\i}{\v{c}}ek}(2013)}]{Zimmermann2013sampling}%
  \BibitemOpen
  \bibfield  {author} {\bibinfo {author} {\bibfnamefont {T.}~\bibnamefont
  {Zimmermann}}\ and\ \bibinfo {author} {\bibfnamefont {J.}~\bibnamefont
  {Van{\'\i}{\v{c}}ek}},\ }\href {\doibase 10.1063/1.4820420} {\bibfield
  {journal} {\bibinfo  {journal} {J.~Chem. Phys.}\ }\textbf {\bibinfo {volume}
  {139}},\ \bibinfo {pages} {104105} (\bibinfo {year} {2013})}\BibitemShut
  {NoStop}%
\bibitem [{\citenamefont {Van{\'\i}{\v{c}}ek}\ \emph
  {et~al.}(2005)\citenamefont {Van{\'\i}{\v{c}}ek}, \citenamefont {Miller},
  \citenamefont {Castillo},\ and\ \citenamefont {Aoiz}}]{Vanicek2005QI}%
  \BibitemOpen
  \bibfield  {author} {\bibinfo {author} {\bibfnamefont {J.}~\bibnamefont
  {Van{\'\i}{\v{c}}ek}}, \bibinfo {author} {\bibfnamefont {W.~H.}\ \bibnamefont
  {Miller}}, \bibinfo {author} {\bibfnamefont {J.~F.}\ \bibnamefont
  {Castillo}}, \ and\ \bibinfo {author} {\bibfnamefont {F.~J.}\ \bibnamefont
  {Aoiz}},\ }\href {\doibase 10.1063/1.1946740} {\bibfield  {journal} {\bibinfo
   {journal} {J.~Chem. Phys.}\ }\textbf {\bibinfo {volume} {123}},\ \bibinfo
  {pages} {054108} (\bibinfo {year} {2005})}\BibitemShut {NoStop}%
\bibitem [{Note4()}]{Note4}%
  \BibitemOpen
  \bibinfo {note} {An important example is RPMD and its TST limit described in
  Refs.~\protect \rev@citealpnum
  {RPMDrefinedRate,rpinst,Hele2013QTST,Althorpe2013QTST}}\BibitemShut {NoStop}%
\bibitem [{\citenamefont {Chandler}\ and\ \citenamefont
  {Wolynes}(1981)}]{Chandler+Wolynes1981}%
  \BibitemOpen
  \bibfield  {author} {\bibinfo {author} {\bibfnamefont {D.}~\bibnamefont
  {Chandler}}\ and\ \bibinfo {author} {\bibfnamefont {P.~G.}\ \bibnamefont
  {Wolynes}},\ }\href {\doibase 10.1063/1.441588} {\bibfield  {journal}
  {\bibinfo  {journal} {J.~Chem. Phys.}\ }\textbf {\bibinfo {volume} {74}},\
  \bibinfo {pages} {4078} (\bibinfo {year} {1981})}\BibitemShut {NoStop}%
\bibitem [{\citenamefont {Alexander}(2001)}]{Alexander2001diabatic}%
  \BibitemOpen
  \bibfield  {author} {\bibinfo {author} {\bibfnamefont {M.~H.}\ \bibnamefont
  {Alexander}},\ }\href {\doibase 10.1016/S0009-2614(01)01012-0} {\bibfield
  {journal} {\bibinfo  {journal} {Chem. Phys. Lett.}\ }\textbf {\bibinfo
  {volume} {347}},\ \bibinfo {pages} {436} (\bibinfo {year}
  {2001})}\BibitemShut {NoStop}%
\bibitem [{\citenamefont
  {Miller}(1975{\natexlab{a}})}]{Miller1975semiclassical}%
  \BibitemOpen
  \bibfield  {author} {\bibinfo {author} {\bibfnamefont {W.~H.}\ \bibnamefont
  {Miller}},\ }\href {\doibase 10.1063/1.430676} {\bibfield  {journal}
  {\bibinfo  {journal} {J.~Chem. Phys.}\ }\textbf {\bibinfo {volume} {62}},\
  \bibinfo {pages} {1899} (\bibinfo {year} {1975}{\natexlab{a}})}\BibitemShut
  {NoStop}%
\bibitem [{\citenamefont {Miller}(1975{\natexlab{b}})}]{Miller1975rate}%
  \BibitemOpen
  \bibfield  {author} {\bibinfo {author} {\bibfnamefont {W.~H.}\ \bibnamefont
  {Miller}},\ }\href {\doibase 10.1063/1.431444} {\bibfield  {journal}
  {\bibinfo  {journal} {J.~Chem. Phys.}\ }\textbf {\bibinfo {volume} {63}},\
  \bibinfo {pages} {1166} (\bibinfo {year} {1975}{\natexlab{b}})}\BibitemShut
  {NoStop}%
\bibitem [{\citenamefont {Lawson}(2000)}]{Lawson2000microcanonical}%
  \BibitemOpen
  \bibfield  {author} {\bibinfo {author} {\bibfnamefont {J.~W.}\ \bibnamefont
  {Lawson}},\ }\href {\doibase 10.1103/PhysRevE.61.61} {\bibfield  {journal}
  {\bibinfo  {journal} {Phys. Rev. E}\ }\textbf {\bibinfo {volume} {61}},\
  \bibinfo {pages} {61} (\bibinfo {year} {2000})}\BibitemShut {NoStop}%
\bibitem [{Note5()}]{Note5}%
  \BibitemOpen
  \bibinfo {note} {In fact this also holds for scattering states bound on the
  product side such that classical trajectories must change direction and come
  back via the crossing point $x=x^\ddagger $.}\BibitemShut {Stop}%
\bibitem [{\citenamefont {Hammes-Schiffer}\ and\ \citenamefont
  {Stuchebrukhov}(2010)}]{HammesSchiffer2010PCET}%
  \BibitemOpen
  \bibfield  {author} {\bibinfo {author} {\bibfnamefont {S.}~\bibnamefont
  {Hammes-Schiffer}}\ and\ \bibinfo {author} {\bibfnamefont {A.~A.}\
  \bibnamefont {Stuchebrukhov}},\ }\href {\doibase 10.1021/cr1001436}
  {\bibfield  {journal} {\bibinfo  {journal} {Chem. Rev.}\ }\textbf {\bibinfo
  {volume} {110}},\ \bibinfo {pages} {6939} (\bibinfo {year}
  {2010})}\BibitemShut {NoStop}%
\bibitem [{\citenamefont {M\"undel}\ and\ \citenamefont
  {Domcke}(1984)}]{Muendel1984morse}%
  \BibitemOpen
  \bibfield  {author} {\bibinfo {author} {\bibfnamefont {C.}~\bibnamefont
  {M\"undel}}\ and\ \bibinfo {author} {\bibfnamefont {W.}~\bibnamefont
  {Domcke}},\ }\href {\doibase 10.1088/0022-3700/17/17/028} {\bibfield
  {journal} {\bibinfo  {journal} {J. Phys. B: At. Mol. Phys.}\ }\textbf
  {\bibinfo {volume} {17}},\ \bibinfo {pages} {3593} (\bibinfo {year}
  {1984})}\BibitemShut {NoStop}%
\bibitem [{\citenamefont {Gradshteyn}\ and\ \citenamefont
  {Ryzhik}(2000)}]{Gradshteyn}%
  \BibitemOpen
  \bibfield  {author} {\bibinfo {author} {\bibfnamefont {I.~S.}\ \bibnamefont
  {Gradshteyn}}\ and\ \bibinfo {author} {\bibfnamefont {I.~M.}\ \bibnamefont
  {Ryzhik}},\ }\href@noop {} {\emph {\bibinfo {title} {Tables of Integrals,
  Series and Products}}},\ \bibinfo {edition} {6th}\ ed.\ (\bibinfo
  {publisher} {Academic Press},\ \bibinfo {address} {San Diego},\ \bibinfo
  {year} {2000})\BibitemShut {NoStop}%
\bibitem [{\citenamefont {Gil}, \citenamefont {Segura},\ and\ \citenamefont
  {Temme}(2004)}]{Gil2004iBessel}%
  \BibitemOpen
  \bibfield  {author} {\bibinfo {author} {\bibfnamefont {A.}~\bibnamefont
  {Gil}}, \bibinfo {author} {\bibfnamefont {J.}~\bibnamefont {Segura}}, \ and\
  \bibinfo {author} {\bibfnamefont {N.~M.}\ \bibnamefont {Temme}},\ }\href
  {\doibase 10.1145/992200.992204} {\bibfield  {journal} {\bibinfo  {journal}
  {ACM T. Math. Software}\ }\textbf {\bibinfo {volume} {30}},\ \bibinfo {pages}
  {159} (\bibinfo {year} {2004})}\BibitemShut {NoStop}%
\bibitem [{\citenamefont {Tannor}(2007)}]{Tannor}%
  \BibitemOpen
  \bibfield  {author} {\bibinfo {author} {\bibfnamefont {D.~J.}\ \bibnamefont
  {Tannor}},\ }\href@noop {} {\emph {\bibinfo {title} {Introduction to Quantum
  Mechanics: {A} Time-Dependent Perspective}}}\ (\bibinfo  {publisher}
  {University Science Books},\ \bibinfo {address} {Sausalito, Calif.},\
  \bibinfo {year} {2007})\BibitemShut {NoStop}%
\end{thebibliography}%

\end{document}